\documentclass[fleqn,10pt]{wlscirep}

\makeatletter
\let\@fnsymbol\@roman
\makeatother

\DeclareMathOperator*{\argmin}{arg\,min}
\DeclareMathOperator*{\argmax}{arg\,max}

\usepackage{placeins}
\usepackage{multirow}
\usepackage{rotating}

\usepackage{changepage}

\title{Heritability maps of human face morphology through large-scale automated three-dimensional phenotyping}

\author[1]{Dimosthenis Tsagkrasoulis}
\author[2]{Pirro Hysi}
\author[2]{Tim Spector}
\author[1,3,*]{Giovanni Montana}
\affil[1]{Department of Mathematics, Imperial College London, SW7 2AZ, London, UK}
\affil[2]{Department of Twin Research and Genetic Epidemiology, King's College London, SE1 7EH, London, UK}
\affil[3]{Department of Biomedical Engineering, King's College London, SE1 7EH, London, UK}
\affil[*]{giovanni.montana@kcl.ac.uk}

\begin{abstract}
The human face is a complex trait under strong genetic control, as evidenced by the striking visual similarity between twins. Nevertheless, heritability estimates of facial traits have often been surprisingly low or difficult to replicate. Furthermore, the construction of facial phenotypes that correspond to naturally perceived facial features remains largely a mystery. We present here a large-scale heritability study of face geometry that aims to address these issues. High-resolution, three-dimensional facial models have been acquired on a cohort of $952$ twins recruited from the TwinsUK registry, and processed through a novel landmarking workflow, GESSA (Geodesic Ensemble Surface Sampling Algorithm). The algorithm places thousands of landmarks throughout the facial surface and automatically establishes point-wise correspondence across faces. These landmarks enabled us to intuitively characterize facial geometry at a fine level of detail through curvature measurements, yielding accurate heritability maps of the human face (www.heritabilitymaps.info).
\end{abstract}
\begin{document}

\flushbottom
\maketitle

\thispagestyle{empty}

\section*{Introduction}

The human face is an important interface of social interaction; communication, sensory input and expression in humans are to a large extent based on facial characteristics and traits \cite{ekman2013emotion}. Normal facial variation is associated with emotional expression \cite{geniole2014fearless}, attractiveness \cite{perrett1994facial} and even lifetime reproductive success \cite{loehr2013facial}. Recent evidence suggest that evolution has contributed to increased diversity and complexity in human facial morphology, presumably due to the role of the face as a primary medium of individual identification and recognition \cite{sheehan2014morphological}. The influence that facial features have in our life has spurred a long and ongoing interest in unraveling the roles that genes and environment play in the morphological characteristics of the human face.

Since the mid-twentieth century, anthropometric scientific research on parent-offspring resemblance and twin concordance has confirmed that variation in human face morphology is driven by genetics \cite{vandenberg1964comparison, lundstrom1988comparison, kohn1991role, hunter1970heritability, nakata1974genetic, lobb1987craniofacial, susanne1977heritability, hauspie1985testing, devor1986transmission}. Heritability studies were carried out to quantify the extent of phenotypic variation that can be explained by genetic variability using, for instance, facial features extracted from cranial measurements. Moderate heritability, varying approximately between $0.35$ and $0.65$, was found for traits such as nasion-basion and nasion-sella distances, as well as the position of the lower jaw and the nasal height \cite{johannsdottir2005heritability, savoye1998genetic, byard1984family, saunders1980family, king1993heritability, harris1991heritability, carson2006maximum, karmakar2007genetic, sherwood2008quantitative}. More recent studies used facial photographs instead, due to the simplicity in which the images can be obtained. However, common traits such as the upper lip height, as well as nasal breadth and vertical eye distance, extracted from standard photographs, were only found to be moderately heritable, with estimates between $0.4$ and $0.53$ \cite{weinberg2013heritability, carels2001quantitative, ermakov2005quantitative, baydacs2007heritability, demayo2010geometric, kim2013heritabilities}. Given the almost perfect resemblance of identical twins, such heritability values appear surprising low. Attempts to replicate these findings across independent studies generated inconsistent evidence. A comparison of eight heritability studies reported low correlation ($<0.4$) between heritability estimates for commonly examined traits such as head circumference, facial height and nose width \cite{kohn1991role}. Further examples include the heritability of facial width, which was reported to range from as low as $0.257$ to $0.629$ \cite{carson2006maximum, devor1986transmission, sparks2002reassessment}, nasal breadth heritability, varying between $0.352$ and $0.639$ \cite{susanne1975genetic, arya2002heritability, devor1986transmission}, and cheek length heritability, ranging from $0.154$ to $0.475$ \cite{carson2006maximum, sjovold1984report}. 

Possible explanations for such unexpectedly low estimates and inconsistency across findings can be found in a number of experimental factors. First of all, in the way in which the traits have historically been measured. Radiographs and photographs are both flat, two-dimensional images. Their use to measure inherently three-dimensional (3D) objects, such as facial surfaces, limits the extent of shape variability that can be captured and constrains the range of facial morphological descriptors that can be extracted. Other issues can be identified in the process of constructing and measuring facial traits. Quantifying face variability is heavily dependent on establishing landmarks across faces. The number of points that can be manually annotated on a face is affected by the type of imaging modality used, and by the ability of a person to establish landmark locations in an unambiguous manner across samples. Consequently, it is common for studies to annotate as landmarks only a few prominent facial markers, such as eye and mouth corners, nose tip and zygomatic bones \cite{johannsdottir2005heritability, savoye1998genetic, byard1984family, saunders1980family, king1993heritability, harris1991heritability, carson2006maximum, karmakar2007genetic, sherwood2008quantitative, weinberg2013heritability, carels2001quantitative, ermakov2005quantitative, baydacs2007heritability ,demayo2010geometric, kim2013heritabilities}, ranging in number between $10$ up to, in exceptional cases, $40$ landmarks. It becomes clear thus that manual landmarking poses an important constraint limiting the extend of facial variability that can be captured. Furthermore, most classical studies adopted facial traits derived from two-dimensional distances between landmark pairs \cite{savoye1998genetic, byard1984family, saunders1980family, harris1991heritability, carson2006maximum, karmakar2007genetic, sherwood2008quantitative,  weinberg2013heritability, carels2001quantitative, demayo2010geometric, kim2013heritabilities}. More rarely, angles between connected landmark pairs have also been considered \cite{johannsdottir2005heritability, king1993heritability, ermakov2005quantitative, baydacs2007heritability}. The widespread adoption of such facial phenotypes could be justified by the small number of annotated landmarks, the relative simplicity in which these measurements can be acquired, as well as their ease of interpretation. On the other hand, they offer a perhaps oversimplified characterization of face morphology and fail to take into full account the geometric variability that can be observed across faces at a more granular level. 

A separate limiting factor that has affected twin heritability studies is related to sample sizes employed and associated statistical modeling implications. Twin studies predominantly employ statistical methods that estimate heritability as the percentage of phenotypic variation that is due to variation of genetic factors \cite{wray2008estimating}. The statistical power of such studies is defined as the probability of correctly rejecting the null hypothesis of zero heritability \cite{westland2010lower}. A multitude of factors affect power, including the combination of variance components used in the model, heritability effect size (ranging between $0$ and $1$), sample size and the proportion of monozygotic (MZ) to dizygotic (DZ) twins in the dataset. How to best optimize the experimental design in terms of sample size is still largely debatable \cite{westland2010lower}. By assuming the existence of only additive genetic effects in the twin model, a previous simulation study reported minimum required sample sizes ranging between $75$ and $2,000$ twin pairs in order to achieve $95\%$ statistical power, depending on heritability effect sizes ranging from $0.8$ to $0.2$ respectively, and keeping the MZ to DZ ratio close to one \cite{martin1978power}. In practice, the large  majority of twin studies to date have relied on sample sizes of $20$ to $100$ twin pairs \cite{vandenberg1964comparison, nakata1974genetic, lobb1987craniofacial, hauspie1985testing, king1993heritability, harris1991heritability, weinberg2013heritability, carels2001quantitative, ermakov2005quantitative, baydacs2007heritability, demayo2010geometric}, possibly due to difficulties in recruiting large cohorts of twins. The use of small to medium sized cohorts may have thus resulted in under-powered studies, especially for traits with low to moderate heritability.

In this work we present a large-scale heritability study of face geometry that departs from previous related investigations in various aspects. First, we acquired 3D facial models. A system for high-resolution 3D photographic scanning, the 3dMD face imaging system, was used to generate anatomically precise three-dimensional polyhedral surfaces of the faces. To capitalize on these representations, we developed a novel automated landmarking procedure, GESSA (Geodesic Ensemble Surface Sampling Algorithm). The algorithm automatically places landmark points throughout the facial surfaces and establishes point-wise correspondence across subjects. GESSA enables the annotation of thousands of landmarks, resulting in the ability to capture morphological variation across subjects at a much finer level of granularity, whilst removing human measurement errors and enabling scalability to large cohorts. The position of each landmark is automatically determined by the algorithm, which attempts to distribute landmark locations uniformly on individual surfaces whilst establishing a precise correspondence across all faces. GESSA was validated on a publicly available dataset of 3D facial surfaces, Morphface \cite{paysan20093d}. The software and validation data are available to download from \url{https://github.com/dimostsag/gessa}. The availability of densely sampled landmark positions on each face enabled a wider range of facial traits to be defined, each capturing a specific aspect of face-shape variability. In this study, we demonstrate that local curvature traits, computed at each three-dimensional position across the facial surface, provide highly informative quantitative measurements of facial geometry, and explore for the first time their heritability. 

A face heritability study was performed on $952$ British twins recruited from the TwinsUK adult twin registry \cite{moayyeri2013cohort}. All subjects were females and unselected for any disease, of which $197$ were MZ and $279$ DZ pairs. To the best of our knowledge, this is the largest twin heritability study of the human face. Each face was represented in the dataset as a 3D polyhedral mesh comprising of approximately $4,500$ points. Using GESSA, we identified $4,096$ landmarks on each face, each one contributing a local curvature value whose heritability was independently assessed. Curvature-based heritability estimates at the individual landmark level were combined into face heritability maps highlighting in great detail, for the very first time, which facial parts are under high and low genetic control. A multivariate analysis involving thousands of closely sampled landmarks further identified extended and well-defined facial regions sharing similar patterns of variability, with heritability estimates reaching or exceeding $0.7$, including the chin, nasal regions, nasolabial folds, upper lips and zygomatic bones. In addition, using a smaller set of these landmarks, we explored the heritability of more traditional distance-based facial traits. A number of facial lengths, including bizygomatic and nose width, had heritability estimates close to or greater than $0.7$, values that are significantly higher than the ones previously encountered in the respective literature. This is the first time that such a detailed and comprehensive evaluation of facial shape heritability has been investigated using a large cohort and 3D data capture technology. Our heritability findings are likely to support future genome-wide studies on facial geometry, while dense representations of facial surfaces through curvature indices may find further use in face recognition and reconstruction techniques. 

\section*{Results}

We present here the phenotyping and heritability results from our sample of $952$ TwinsUK twins. Point correspondences for $4,096$ landmarks were automatically established using GESSA. Shape-related phenotypes were constructed using four different curvature indices on the landmark sets. Visualization of heritability estimates associated to landmark-wise curvature traits produced high-definition heritability maps. A multivariate analysis of these landmark-wise measurements, based on sparse PCA (sPCA), indicated the presence of spatially coherent traits extending over larger areas of the face, whose heritability was also estimated. Finally, a subset of $17$ landmarks was selected and the heritability of $20$ traits based on Euclidean and Geodesic distances was calculated.

\subsection*{Curvature-based Morphological Traits in the TwinksUK dataset}

\begin{figure}[ht]
\centerline{\includegraphics[width=4.0in]{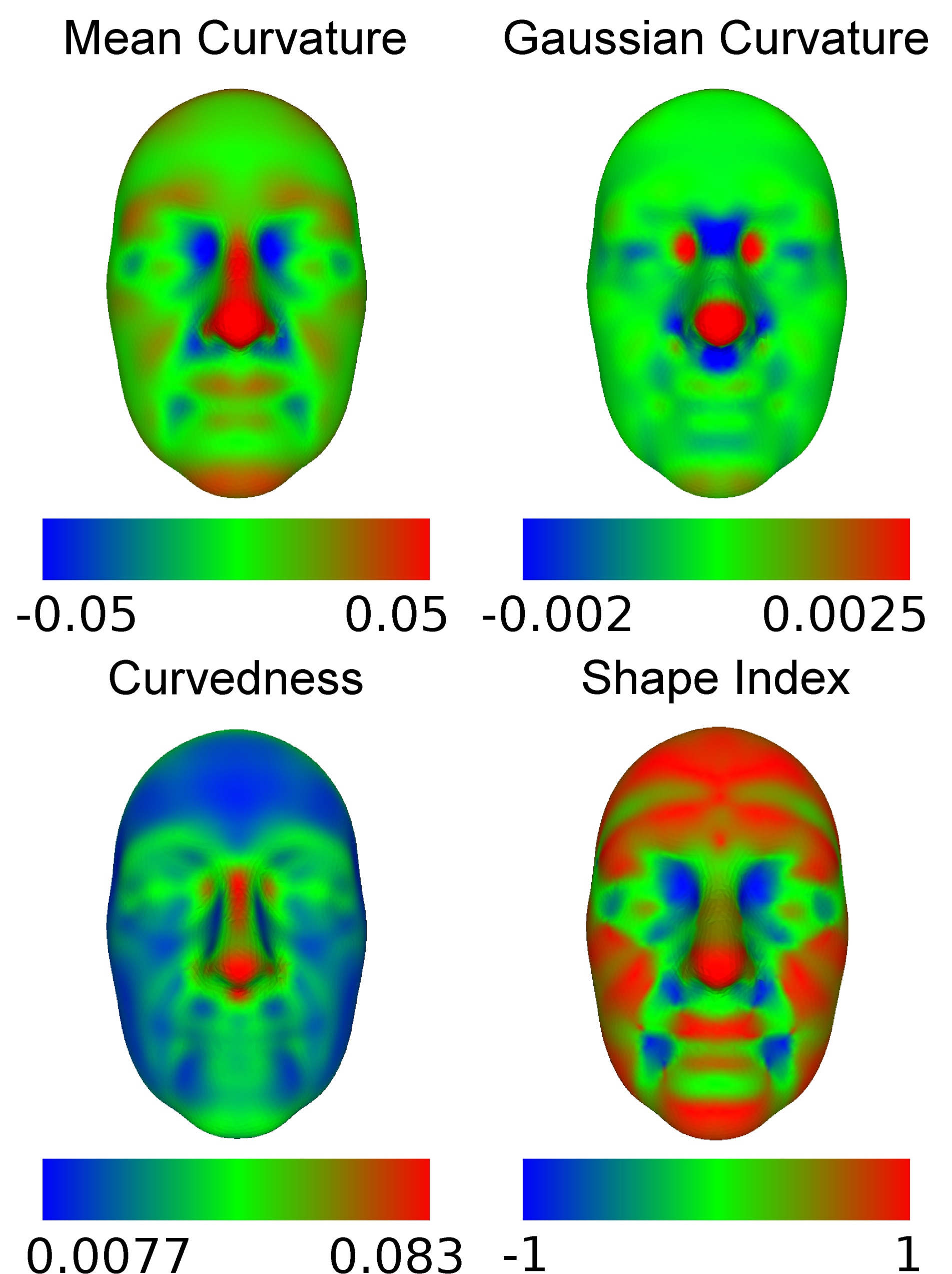}}
\caption{{\bf Face curvature maps. } The figure shows the curvature maps of the average face from the TwinksUK dataset. The maps were created by computing values of curvature indices along all landmarks on the face. Most facial parts have a convex topology, either cylindrical or spherical, with few saddle-like or flat areas that transition to concave regions localized in the eye, nose and mouth corners. Each map further highlighted the unique characteristics of its corresponding index.
{\bf Mean Curvature Map:} Positive values corresponded to convex areas (nose, eyebrows, lips, chin), while negative values to concave ones (inner eye corners, subnasal region). 
{\bf Gaussian Curvature Map:} Many facial regions had very small values, due to being either flat or curved along only one direction. Exceptions were the spherical (nose tip, inner eye corners) and saddle-like (nasion, subnasal region) areas. 
{\bf Curvedness Map:} Consisting only of non-negative values, the map highlighted how curved the surface is without distinguishing between different shape morphologies. Large values were concentrated on the central part of the face. 
{\bf Shape Index Map:} Most facial areas had positive values, indicating the overall cylindrical structure of the human face. 
}
\label{fig_1}
\end{figure}

The set of $952$ three-dimensional facial surfaces from the TwinsUK cohort was processed with GESSA and $4,096$ landmarks were identified. Each landmark contributed four shape-related phenotypes, associated with how bent the surface is around that point. These traits were computed using local curvature indices, namely Mean Curvature (MC), Gaussian Curvature (GC), Curvedness (CU) and Shape Index (SI), resulting in $16,384$ quantitative traits per face. A detailed description of the curvature indices, and the rationale for using different types, can be found in the Methods section. For each one of these four measures, curvature maps of the average TwinsUK face - constructed by averaging landmark positions of all faces - were obtained by color-coding all facial landmarks according to their curvature values. Figure \ref{fig_1} shows the resulting maps, which provide easily interpretable representations of facial morphology and underline the different attributes of each curvature index. It must be noted that, due to ethical reasons, individual faces cannot be shown. We only visualize the average face, computed by averaging landmark coordinates of the $952$ faces. Due to the large sample size of our dataset, the average face appears somewhat smoothed, in particular around the mouth and eye regions. Supplementary Figure 10 shows average faces obtained using only $10$ and $200$ individuals. It can be clearly noticed that the increasing sample size has a smoothing effect on the average face.

The MC index provides a balanced measure between shape morphology, i.e. flat vs. cylindrical vs saddle structure, and curvature magnitude, i.e. how bent the surface is irrespective of shape. In the MC curvature map, points belonging to protruding concave regions like the nose, chin and eyebrows had positive MC values, with higher measurements observed in the nasal surface. Flatter areas, such as cheeks and forehead, had MC values close to zero, while inner eye corners, ala of the nose, and to a lesser extent, the corner areas between lips and chin were comprised of negative-valued curvatures. GC describes well variation between and within cylindrical and saddle-like structures, while being less sensitive to other shape characteristics. In the GC map, facial features associated with positive values were the cylindrically structured inner eye corners and nose tip, while saddle-like regions such as nasion and base of the nose showed negative GC values. CU is affected by changes in the magnitude of the curvature but not by shape morphology. The CU curvature map highlighted facial parts with large overall curvature, for example nose and eyebrows, which confirms the intuitive observation that highly curved facial parts are mainly centrally located in the face. Contrary to CU, the SI index primarily distinguishes between different shape morphologies, but is less sensitive to curvature magnitude. The SI curvature map showed that that most facial areas have positive values corresponding to generally cylindrical structures.

The sample variability of each curvature trait was also investigated. Curvature variance maps can be found in the Supplementary Fig. 4. 
Depending on the curvature index, various facial areas showed increased variability. High variance of the MC measurements was observed in areas such as eye sockets, ala of the nose and mouth. GC and CU variance was located mainly in the nose, eye and philtrum regions, while the SI captured increased variation in the zygomatic and mouth areas. It is important to notice that particular facial areas, namely the mouth and eye regions, showed consistently high variability, which could relate to the increased motional ability of the specific structures. Irrespective of the curvature index used, average phenotypic variability was always higher in the subset of DZ, compared to the MZ pairs. As heritability is based on differences in similarity between MZ and DZ pairs \cite{boomsma2002classical}, we also computed the means and variances of absolute trait differences between pairs. The results were combined into facial maps (Supplementary Fig. 5) 
and showed clearly higher values in the DZ subset, compared to the MZ one. 

\subsection*{Univariate Heritability Analyses}

We performed univariate analyses of all $16,384$ curvature traits - $4$ traits per landmark - with the aim to combine local heritability results and produce global maps of heritability for the human face. The heritability of each trait was independently estimated using Structural Equation Modeling (SEM) \cite{rijsdijk2002analytic}. The method evaluates which combination of additive (A) genetic, common (C) environmental and unique (E) environmental variance components can best explain the observed phenotypic variance and covariance of MZ and DZ twin data; see Methods for a detailed description of the  model approach. Different combinations of A, C and E component models were considered. The Akaike Information Criterion (AIC) \cite{akaike1974new} was used to guide model selection. AE models were the best-fitting ones according to AIC. Summary statistics for all fitted models can be found in Supplementary Table 1. 
SEM also assessed the ability of the model to fit the observed data. Model Goodness-of-fit was examined using a log-likelihood ratio test between the structured model and a fitted saturated model where no structure was imposed on the covariances. Test $p$-values above $0.05$ translated to the structured model providing a better fit than the saturated one. Details regarding goodness-of-fit can be found in the Supplementary Text. From the sets of $4,096$ traits per index, $86.02\%$, $67.9\%$, $86.8\%$ and $87.06\%$ AE models for MC, GC, CU and SI respectively had goodness-of-fit $p$-values above $0.05$.

For each curvature type, the $4,096$ heritability estimates were visualized in a single heritability map plot. Each map provides a graphical representation of the extent by which the geometry of facial regions is controlled by genetic variability. Frontal and side facial views of these maps are shown in Fig. \ref{fig_2}. For interactive viewing of the heritability maps, a website was created at \url{http://heritabilitymaps.info/}.

Several landmarks with high ($> 0.65$)  heritability estimates were localized on well defined facial areas. Irrespective of the curvature index employed, landmarks belonging to the mental region, philtrum, nasal tip, nasion, inner eye corners, nasolabial folds and frontal process of maxilla gave consistently high estimates. Amongst all curvature types, MC yielded the highest heritability estimates over extended facial regions. The MC heritability map highlighted further highly heritable areas, including the zygomatic lines around the eye sockets, side areas of the mental foramen, the upper lip and frontal eminences.  The results of the GC traits showed high heritability for a number of saddle-like facial structures, namely the whole nasion region, the philtral ridge, as well as the lower nasal bone. In the CU heritability map, a strong genetic influence is observed in the angular transition from the nasal bone and glabella towards the frontal eminences, as well as in the overall roundness of the facial circumference, highlighted by heritable lines across the upper part of the forehead and the lower part of the ramus of the mandible.  The SI heritability map showed strong genetic control of the softly spherical flat regions of frontal eminences and upper lip, the cylindrical structure of the upper zygomatic bones and sides of the nose and finally the saddle-like areas of nasion, nasal bone and ala of the nose. The results indicated that local curvature is strongly determined by genes for large parts of the face.

\subsection*{Multivariate Heritability Analyses}

We aimed to identify larger facial areas showing common patterns of shape variation and produce single-valued heritability estimates for them. For each face and curvature index, we decomposed the aggregated landmark phenotypes using sparse Principal Component Analysis (sPCA) \cite{witten2013package, witten2009penalized} (see Supplementary Text for further details on sPCA). sPCA automatically identified traits comprised of linear combinations of landmark phenotypes with similar curvature variability. We refer to these as regional traits. Sparsity affected the number of landmark traits comprising each regional phenotype, and was imposed in order to acquire measurements corresponding to extended but spatially consistent facial regions. The amount of sparsity was controlled through a single parameter. Different values were tested and results showed that the parameter had little effect in the heritability estimation process (Supplementary Fig. 3). 
The parameter was set to $7.5$ for GC, $12.5$ for SI and $15$ for MC and CU. We estimated the heritability of $100$ regional traits for each curvature index. Each set of regional traits captured approximately $90\%$ of its corresponding curvature's sample variability. The cumulative amounts of explained variances for all regional traits are reported in the Supplementary Files $1-4$, for MC, GC, CU and SI indices respectively.

Average heritability statistics for the regional trait analyses are included in Supplementary Table 1. 
We found again that the best models were the AE as assessed by AIC. Goodness-of-fit $p$-values above $0.05$ were acquired for $81$, $75$, $87$ and $73$ regional traits of MC, GC, CU and SI respectively.

The facial areas associated to regional traits were visualized by color-mapping the coefficients of the linear combinations - weights by which landmark traits contribute to the regional phenotypes - on the facial surface. We refer to these maps as Eigenface maps, since they correspond to the eigenvectors of the multivariate decomposition. Eigenface maps of the $10$ highest variance explaining regional traits for each curvature index are shown in Supplementary Fig. 2. 
Figure \ref{fig_3} shows the Eigenface maps of the top $5$ heritable regional traits for each curvature index, along with their heritability values. Corresponding phenotypic correlations for MZ and DZ subsets are reported in Supplementary Table 2. 
The maps of the most heritable traits were mostly comprised of landmarks closely located to each other, indicating patterns of common shape variation in the respective areas. 

Heritability results for the regional and univariate traits were in good agreement. From Fig. \ref{fig_3} it is obvious that the morphologies of regions such as nasolabial folds, zygomatic bones, inner eye corners, mental region, frontal eminences and ala of the nose were highly heritable ($> 0.65$). Due to the good segmentation of the facial surface into clearly identifiable regional traits, we also identified heritable areas that were not as easily noticeable in the heritability maps. In particular, the mental foramen showed up as the second top heritable phenotype in the SI results, while the condyloid process of the mandible was highlighted in the fifth and third top heritable trait in MC and SI analyses, respectively. In the discussion, we identify regional traits by their curvature index and their variance-explaining order, as shown in Fig. \ref{fig_3}.

\subsection*{Heritability of Distance-based Traits}

An analysis of traditional distance-based facial traits was performed to gain insights about the relative merits of our phenotypes and also enable direct comparisons to previous heritability studies. Out of the $4,096$ landmarks, we located $17$ corresponding to prominent fiducial points, by visual inspection of the average TwinsUK face. Figure \ref{fig_4} shows the selected landmarks. Using the computed correspondence, we were able to automatically locate the $17$ landmarks on all $952$ faces. Ten facial traits derived from Euclidean distances (EDTs) between selected landmark pairs were subsequently considered. The phenotypes are summarized in Table \ref{table_1}. In addition, we constructed ten equivalent distance traits measured as lengths of connecting paths between the same landmark pairs, where the paths were restricted to lie only on the facial surfaces. Such distances, defined on non-flat surfaces, are called Geodesic distances. An illustration of the difference between the two types of distances can be seen in Supplementary Fig. 6. 
We examined whether the use of facial traits derived from Geodesic distances (GDTs), only possible on 3D data, yielded any advantages compared to Euclidean traits.

Heritability estimation for the $10$ EDTs and their equivalent GDTs was carried out using ACE, AE and E structural equation models, as before. AE models provided on average the best fits according to AIC. Supplementary Table 3 
provides detailed statistics for the fitted models. Table \ref{table_2} shows the resulting heritability estimates as well as SEM Goodness-Of-Fit test $p$-values. In the remainder, we concentrate on traits whose models' Goodness-Of-Fit test $p$-values were greater than $0.05$, thus providing good fits of the observed data. Four EDTs corresponding to horizontal facial measurements, namely nose, zygomatic, mandible and mouth widths were found to be highly heritable. Of these, estimates greater than $0.7$ were acquired for the two upper/middle face EDTs, zygomatic and nose widths, while the heritabilities of the mandible and mouth widths were found to be slightly lower at approximately $0.62$ and $0.67$, respectively. The EDT corresponding to nasal protrusion was also found to be moderately heritable at approximately $0.55$. Heritable GDTs included the mandible and mouth widths, nasal protrusion, lower face height and biocular width (i.e. the distance between inner eye corners). The first three GDTs had slightly lower heritability estimates compared to the corresponding EDTs.  Mandible and mouth GDTs traverse the lip region, which had low heritability estimates in our previous analyses. On the other hand, the biocular GDT, which traverses a surface area with consistently high heritability, i.e the nasion, provided the highest estimate ($0.789$) among all distance phenotypes.

\section*{Discussion}

We presented a novel landmarking and phenotyping methodology for 3D surfaces and performed a large-scale twin heritability study of the human face. A salient aspect of our analysis is the automated dense landmarking procedure. Dense landmarking approaches have been recently adopted in face modeling and candidate association analyses in order to study genetic syndromes involving facial dysmorphisms and asymmetries \cite{hammond2012large, claes2012improved} and recognize genetic variants that explain normal face variation \cite{peng2013detecting, claes2014modeling}. Existing methods, though, are either still heavily dependent on some form of manual landmarking, which can be a tedious and error-prone process, or not suited for the analysis of polyhedral surfaces. Further details on related work can be found in the Supplementary Text. Here, we propose a new methodology, GESSA, which makes use of appropriate mathematical structures, such as distances and paths, directly defined on the surfaces, in order to provide uniform and dense landmarking of 3D polyhedral models in an accurate and efficient manner. The GESSA software and validation data can be freely downloaded from \url{https://github.com/dimostsag/gessa}. Non sensitive data, such as curvature data matrices, are available upon request to the authors. Furthermore, this is the first time that dense facial landmarking has been used in a twin heritability study. 

A second novel aspect of our methodology, facilitated by the three-dimensional face models, is the use of curvature-based phenotypes. For each landmark point in the surface, four different types of univariate measurements describing curvature for local patches centered around the landmarks were considered. Each curvature type is able to highlight varying morphological structures. The use of these traits enabled us to characterize local variability in facial shape and identify its genetic content. Heritability estimates for individual landmark traits were combined to provide detailed global maps of heritability for the human face. Furthermore, regional traits, defined as linear combinations of the single landmark traits, were computed by a multivariate decomposition of the previous traits. Heritability analyses of the latter phenotypes allowed us to report on accurate heritability values for well-defined facial regions. In the literature, similar curvature traits have been successfully used for other applications such as face detection \cite{colombo20063d}, recognition \cite{inan20123, dorai1997cosmos}, segmentation \cite{salazar20103d, ceron2010relevance} and affinity estimation \cite{ceron2010relevance, zhao20143d}.

Our landmarking and phenotyping pipeline was employed for the analysis of $456$ female twin pairs from the TwinsUK cohort \cite{moayyeri2013cohort}. Previous work has suggested that the TwinsUK sample is representative of the general British population, where the sample were ascertained from \cite{andrew2001twins}. For facial traits in particular, it may be expected that gender also plays a role in shaping up the facial characteristics we study, such as curvature measures. For this reason, caution is warranted before any of these results are generalized and extended to male subjects. 

To our knowledge, this is the largest face heritability study ever done. Compared to most commonly encountered sample sizes of between $20$ and $100$ pairs \cite{vandenberg1964comparison, nakata1974genetic, lobb1987craniofacial, hauspie1985testing, king1993heritability, harris1991heritability, weinberg2013heritability, carels2001quantitative, ermakov2005quantitative, baydacs2007heritability, demayo2010geometric}, the increased number of subjects improves the statistical power of our study to identify effects of heritability. Based on a previously published simulation study on the power of twin studies \cite{martin1978power}, our sample size surpasses the minimum requirements for having $95\%$ power of rejection of the false - zero heritability - hypothesis at the $5\%$ significance level even when the true heritability effect is as low as $0.3$.

\begin{figure}[ht]
\centerline{\includegraphics[width=4.5in]{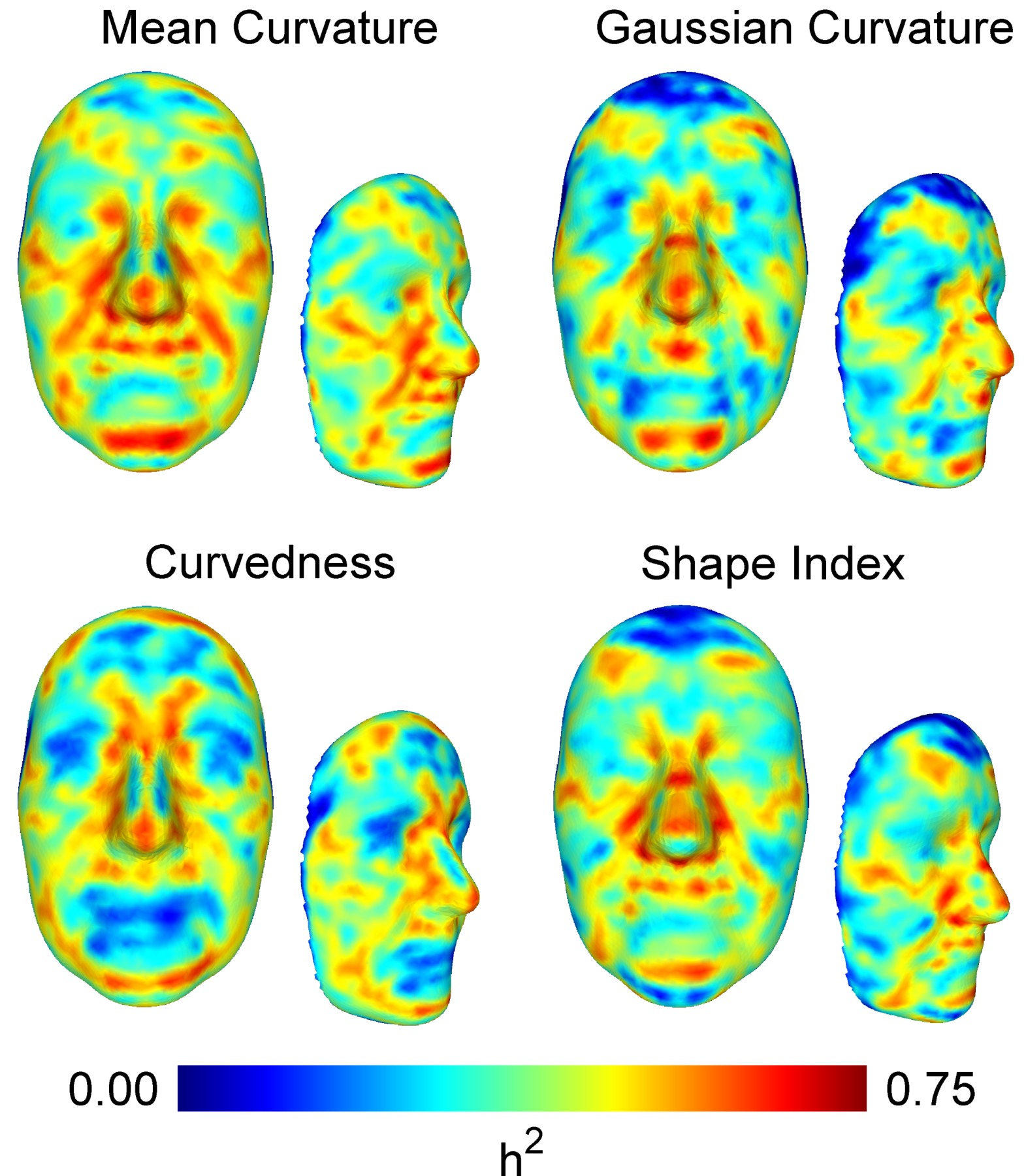}}
\caption{{\bf Heritability maps of the human face.} Each heritability map consists of $4,096$ landmark heritability estimates. Facial areas with high heritability across all four maps were the mental region, philtrum, nasal tip, nasion, inner eye corners, nasolabial folds and frontal process of maxilla. 
{\bf Mean Curvature Heritability Map:} In addition to aforementioned areas, we observed genetic association along the lower parts of the eyes (zygomatic bones), the sides of mouth/chin complex (mental foramen), the complete upper lip region and the frontal eminences. 
{\bf Gaussian Curvature Heritability Map:} Compared to the previous map, zygomatic bones were not clearly distinguished and only the philtrum had high heritability in the upper lip region. A moderate genetic effect was observed around the whole nasion area. 
{\bf Curvedness Heritability Map:} The map included clear heritable lines in the upper and lower circumference of the facial surface. A further unique feature observed here was a flat heritable area around the ramus of the mandible. 
{\bf Shape Index Heritability Map:} Zygomatic lines were clearly visible in this map, along with the upper lip and the frontal eminences. The map highlighted primarily the mostly concave regions of transition between the lip and ala of the nose, the frontal process of maxilla and the lower more protruding parts of the nasal bones.
}
\label{fig_2}
\end{figure}

\begin{figure}[ht]
\centerline{\includegraphics[width=5.3in]{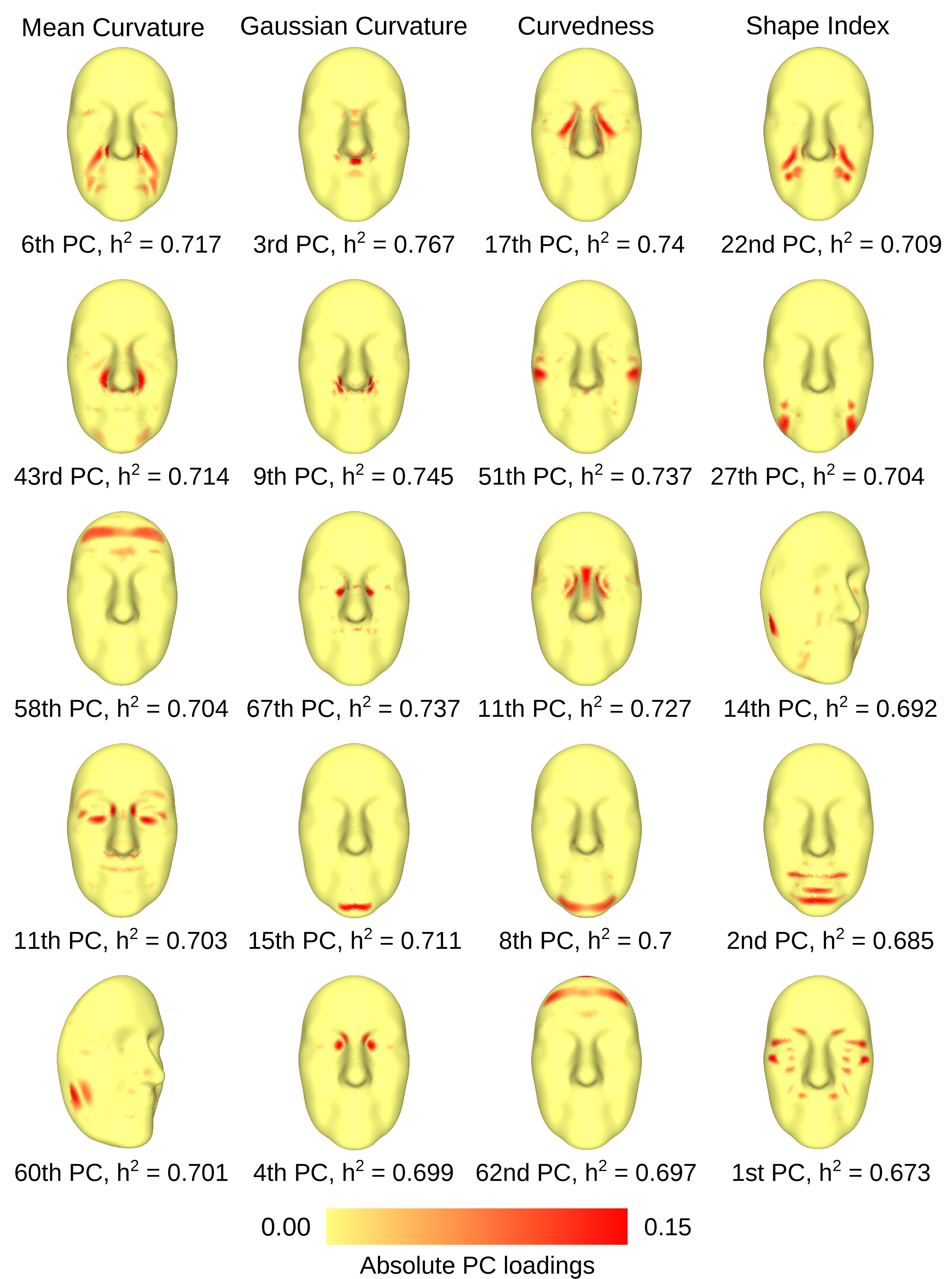}}
\caption{{\bf Eigenface maps of the top heritable regional traits.} The maps depict the weights by which landmark phenotypes contribute to the regional traits. sPCA was utilized in order to achieve good spatial consistency of the Eigenface maps.
{\bf Mean Curvature Eigenface Maps:} Distinctive facial characteristics that emerged as heritable were the nasolabial folds, transitions to the ala of the nose, frontal eminences, zygomatic areas between and below the eye sockets and the condyloid process of the mandible, with the latter been more clearly portrayed in the multivariate rather than the univariate heritability analysis. 
{\bf Gaussian Curvature Eigenface Maps:} The GC Eigenface maps were highly localized, compared to the rest of the curvature indices. Heritable regions included for the philtrum, ala transitions, inner eye corners and chin facial regions.
{\bf Curvedness Eigenface Maps:} The top heritable traits highlighted mainly the zygomatic areas around the eye sockets, as well as the nasion, chin and upper forehead areas.
{\bf Shape Index Eigenface Maps:} Highly heritable regional traits were located in the nasolabial folds, zygomatic areas,chin, condyloid process of the mandible and mental foramen regions. 
}
\label{fig_3}
\end{figure}

\begin{figure}[ht]
\centerline{\includegraphics[width=2.5in]{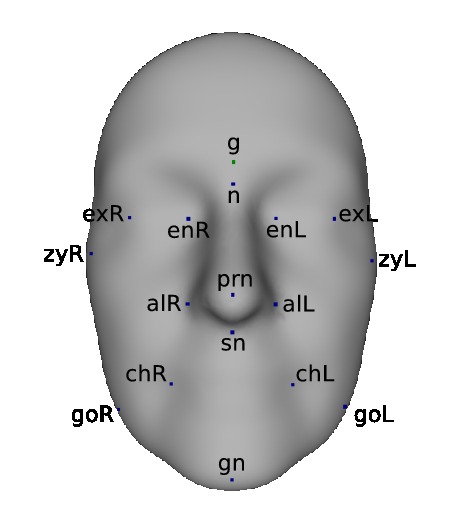}}
\caption{{\bf Distance-based analysis - Selected landmarks depicted on the average TwinsUK facial surface. } $17$ facial points corresponding to prominent fiducial markers were selected from the automatically computed set of $4,096$ landmark points. Landmark pairs were subsequently used to construct distance-based phenotypes. See Table \ref{table_1} for naming conventions.}
\label{fig_4}
\end{figure}

We were able to identify curvature-based facial traits that were highly heritable ($> 0.65$) in both of our curvature-based analyses. A direct comparison with previously reported heritable facial lengths and angles is not straightforward, due to the different nature of the measurements. Certain connections though were made between our heritability and Eigenface maps and related published findings. A heritability estimate of $0.53$ was previously reported for nose width in a pedigree analysis of $229$ Korean individuals \cite{kim2013heritabilities}. The MC heritability map (Fig. \ref{fig_2}) showed high heritability for the line between the left and right corners of the ala through the base of the nose. In the same study, inner eye corner distance had a heritability estimate of $0.61$. The GC, CU and SI heritability maps all indicated that the shape of the nasion region, including the inner eye corners, was highly heritable. This inference was further supported by our multivariate analysis. The $3rd$ and $5th$ most heritable GC regional traits, as well as the $3rd$ top heritable CU trait (Fig. \ref{fig_3}) had Eigenface maps concentrated on the nasion area (Fig. \ref{fig_3}) and their respective heritability estimates ranged between $0.699$ and $0.737$. Moderate to high heritability, ranging between $0.4$ and $0.7$, was identified in twin and family studies for the facial width \cite{baydacs2007heritability, sherwood2008quantitative, ermakov2005quantitative, karmakar2007genetic}. Our MC and SI heritability maps (Fig. \ref{fig_2}) highlighted highly heritable lines following the curve of the zygomatic bones. Another phenotype that was reported with heritability estimates of $0.59$ and $0.66$ is head circumference \cite{ermakov2010family, karmakar2007genetic}. This result can be connected to the elevated heritability estimates regarding curvature magnitude in the periphery of the face, that we observed in the CU heritability map (Fig. \ref{fig_2}). Finally, a number of twin and family studies identified moderate to high heritability estimates for various phenotypic traits relating to the position, length and angular structure of the jaw bone. Mandible ramus and body length heritabilities were reported to be $0.72$ and $0.77$ respectively in a study of $363$ children and their parents \cite{johannsdottir2005heritability}, while the angle between the two lines was found to have moderate heritability $0.47$ and $0.453$ in the same analysis and in a different study of $77$ twins \cite{carels2001quantitative}. Chin width was estimated to have heritability of $0.42$ in a family study \cite{kim2013heritabilities}. In our multivariate results, we observed high estimates in the chin area - $2nd$ SI, $8th$ CU and $15th$ GC regional traits, with respective trait heritabilities ranging from $0.685$ to $0.711$ (Fig. \ref{fig_3}). All above comparisons are summarized in Supplementary Table 4. 

We also explored the heritability of facial traits based on Euclidean distances (EDTs), as well as equivalent phenotypes measured using Geodesic distances on the facial surfaces (GDTs). $10$ EDTs and GDTs yielded reliable heritability estimates, ranging from $0.505$ to $0.789$. Evidence supporting our results were found in a number of family-based heritability analyses. A previously mentioned study reported its highest heritability estimates of $0.42$, $0.44$, $0.53$ and $0.61$ for mandible width, nasal protrusion, nose width and inner eye corner distance respectively \cite{kim2013heritabilities}. Here, the nose width EDT and biocular width GDT yielded two of the three highest estimates -$0.718$ and $0.789$ respectively-, while nasal protrusion and eye distance EDTs and GDTs were moderately to highly heritable, with heritability estimates ranging from $0.505$ to $0.677$. Nose width was also reported to be moderately to highly heritable in two studies of $125$ Belgian and $342$ Indian families with corresponding estimates of $0.639$ and $0.498$ \cite{susanne1977heritability, arya2002heritability}. The same studies reported heritabilities of $0.606$ and $0.605$ for the bizygomatic width, while three further analyses, including large population samples from Russia and India and Europe, gave estimates of $0.52$, $0.71$ and $0.629$ for the same trait \cite{ermakov2005quantitative, karmakar2007genetic, sparks2002reassessment}. Our most heritable EDT, with $h^2 = 0.734$ corresponded to that width. In comparison to previously reported values, our heritability estimates were either similar or significantly higher.

Important inferences were made by exploring results from the various types of phenotypes used in this study. In the distance-based study, EDTs corresponding to mandible and mouth widths showed slightly higher heritability estimates in comparison to the respective GDTs. The latter traits had geodesic paths passing through the lip region, which, in the curvature analyses (Fig. \ref{fig_2}), had consistently low heritability estimates. On the other hand, the GDT corresponding to inner eye corner distance gave the highest heritability estimate of $0.789$. The geodesic path of that trait traverses the nasion region, which is a well known highly heritable area of the human face \cite{paternoster2012genome, kim2013heritabilities, weinberg2013heritability}. The results could indicate that GDTs are more sensitive than EDTs to the facial morphology between the considered landmarks. A comparison of summary statistics from the curvature and distance heritability analyses revealed that Structural Equation Models provided good data fits for approximately $80\%$ of the curvature traits, but only $50\%$ of the distance phenotypes, as assessed by Goodness-Of-Fit log-likelihood ratio tests. This observation could be a strong indicator that curvature phenotypes are more suitable for the study of facial morphology. 

\section*{Conclusion}

In this work we proposed a new method for the automated and dense landmarking of 3D surfaces, GESSA, and applied it in a novel large-scale twin heritability study of the human face morphology. Heritability estimates were computed for local curvature phenotypes corresponding to single landmarks and the results were combined to generate face heritability maps. Furthermore, regional curvature traits, corresponding to larger facial areas, were extracted through a multivariate analysis and their heritability was also estimated, yielding a number of highly heritable facial features. Heritability estimation was also performed for a number of traditional facial length traits, with estimates being equivalent or higher than the ones found in the existing literature. In conclusion, we provided a fresh perceptive in facial phenotyping and heritability analysis that could potentially inform future genome-wide studies and be useful in a variety of applications, ranging from population genetics and gene-mapping studies, to face modeling and reconstruction applications.

\section*{Methods}

\subsection*{Sample Description}

Imaging data were retrieved for $1,547$ participants of British origin from the TwinsUK Cohort \cite{moayyeri2013cohort}, for which 3D facial models and age information was available. All subjects provided written and informed consent for academic use of the data. Experiments were approved by the Guy’s and St. Thomas’ (GSTT) Ethics Committee. The research was done in accordance with the tenets of the Declaration of Helsinki. The sample consists predominately of female twins unselected for any disease. Detailed sample characteristics of the TwinsUK Cohort have been previously described \cite{moayyeri2013cohort}. $314$ subjects were dropped due to artifacts in the images or lack of zygosity information. $228$ unrelated subjects were excluded from further analysis. An additional $56$ subjects were removed due to mesh reconstruction errors during our phenotyping process. After preprocessing and quality control, we were left with $952$ female subjects ($59.31$ mean age, standard deviation $9.85$). Of these, $197$ pairs were monozygotic and $279$ were dizygotic.

\subsection*{Image Acquisition and Preprocessing}

The raw 3D facial images were acquired using a 3D photographic scanning system manufactured by 3dMD. Participants were asked to keep their mouths closed and a neutral expression during the acquisition of the 3D scans. Each image was comprised of a 3D triangular mesh, with approximately $4,500$ points representing the frontal facial surface, and the corresponding texture map. Texture maps were not used in our analysis. Due to large variation in the original pose and position of the original meshes, we manually located outer eye corners and nasion in all faces. The three landmarks were used to impose a common orientation of the faces under the same coordinate space. High landmark accuracy was not important for this purpose and these landmarks were discarded from any subsequent analysis step. Following that, the surfaces were cropped and trimmed to remove non-facial areas, such as neck and chest regions, hair and ears. Finally, the Iterative Closest Point (ICP) algorithm \cite{besl1992method} was applied to align the cropped images. The preprocessing pipeline was performed using the Meshlab software suite.

\subsection*{The Geodesic Ensemble Surface Sampling Algorithm (GESSA)}

For the streamlined identification and alignment of landmark points across 3D faces, we propose and use our novel landmark sampling procedure, Geodesic Ensemble Surface Sampling Algorithm (GESSA). GESSA automatically samples large sets of corresponding landmark points from sets of similar polyhedral surfaces. We formulate the problem of finding corresponding landmarks as a minimization of an objective function comprised of two entropy-based terms. The first term is the entropy of the data probability distribution. Minimization of this term is performed by moving landmark positions and leads to improvement of landmark correspondences across surfaces. The second term is the sum of entropies for landmark distributions on individual surfaces. By maximizing this term the algorithm achieves a uniform distribution of points on all related surfaces. 

The above formulation is adopted from the work by Cates et. al \cite{cates2007shape}, which was, in turn, based upon previous work on statistical shape modeling with information-based optimization functions. The first such work that we are aware of was presented by Kotcheff and Taylor \cite{kotcheff1998automatic}, and subsequent articles by Davies et al. \cite{davies2002minimum, davies20023d}. Further details on these methods can be found in the Supplementary Text. Here we built upon that framework and use geodesic distances between landmarks, directly defined on the polyhedral facial surfaces, in order to increase precision during uniform landmark sampling. Furthermore, a suitable gradient descent optimization technique was developed to optimize landmark locations by operating only on the surface structures. By incorporating these two key features, we were able to to deal with highly curved surfaces, improve upon computational space requirements and enhance the correspondence results.

\subsubsection*{General Methodology}

Let us consider an ensemble of $N$ polyhedral surfaces. Each surface $\mathcal{S}_j$, $j=1 \ldots N$, is represented, without loss of generality, as a a triangular mesh in $R^3$. Our objective is to sample $M$ points $x_j^1, \ldots, x_j^M$, with $x_j^k = ({x_j^k}_1, {x_j^k}_2, {x_j^k}_3) \in \mathcal{S}_j$, uniformly from each surface, and establish one-to-one correspondence among points on all surfaces. As such, the overall correspondence problem can be broken down to two major components; correspondence optimization of landmarks across the ensemble and uniform sampling on individual surfaces.

The coordinates of $M$ points on a surface $\mathcal{S}_j$ can be aggregated into a vector $z_j$, with
\begin{equation}
z_j = ({x_j^1}_1, {x_j^1}_2, {x_j^1}_3, \ldots, {x_j^M}_1, {x_j^M}_2, {x_j^M}_3),
\end{equation}
$z_j \in R^{3 M}$. The vectors $z_j$ can be thought of as point representations of surfaces distributed in $R^{3 M}$. As such, $R^{3 M}$ is taken to be the space of all surfaces, when each one is sampled in $M$ locations, referred hereafter as shape space.  Consider $Z \in R^{3 M}$ to be a random variable in shape space with probability density function $p(Z)$, and $z_j$, $j=1 \ldots N$, realizations of that random variable. The differential entropy of $Z$ is given by  $\Gamma(Z) = E[- \log p(z)] = - \int_{\mathcal{Z}} p(z) \log p(z) dz$. As such, the sample differential entropy is given by
\begin{equation}
\bar{\Gamma}(z_1, \ldots, z_N) = - \frac{1}{N} \sum_{j=1}^N \log p(z_j). 
\end{equation}

One-to-one correspondence of landmarks can be optimized by minimizing the above sample entropy. This minimization increases the compactness of the surfaces' distribution in shape space, which equates to bringing surface landmarks closer to each other. A potential risk though is that landmarks can be collapsed to the same surface locations. The solution is to balance the correspondence accuracy with uniform distributions of points on individual surfaces.

Assuming that $x_j^1, \ldots, x_j^M$, have been sampled in some way from the surface $S_j$, their positions can be manipulated in order to make them uniformly distributed on the surface. This is done by maximizing the sample entropy for the distribution of landmarks in $S_j$, since, in bounded domains, as are our surfaces, the uniform distribution has maximum entropy. Let $X_j \in \mathcal{S}_j$ be a random variable on surface $\mathcal{S}_j$ with probability density function $p_j(X_j)$, and  $x_j^1, \ldots, x_j^M$, realizations of $X_j$. The differential entropy of $X_j$ is $H_j(X_j) = E[-\ln p_j(x)] = - \int_{\mathcal{S}_j} p_j(x) \ln p_j(x) \partial x$, and the sample differential entropy is given by
\begin{equation}
\bar{H}_j(x_j^1, \ldots , x_j^M) = - \frac{1}{M} \sum_{k=1}^M \ln p_j(x_k) .
\end{equation}

The combined optimization cost used in the correspondence algorithm balances the sample entropy in shape space with the sum of point distribution entropies and is given by
\begin{equation}\label{combined}
Q = \bar{\Gamma}(z_1, \ldots, z_N) - \sum_{j=1}^N \bar{H}_j(x_j^1, \ldots , x_j^M),
\end{equation}

$Q$ must be minimized under a set of constraints imposing that each point lies in its corresponding surface. We propose the use of a geodesic gradient descent algorithm which directly follows straightest paths on the surfaces in order to update landmark locations without violating these constraints. We first introduce some notions related to manifold geometry that are required to develop our methodology. We then describe the details regarding uniform distribution of points on individual surfaces, followed by the correspondence optimization.

\subsubsection*{Geometric Structures on Manifolds}

The structure of a 3D object is commonly described by its boundary surface in $R^3$. Such surfaces can be mathematically described as 2D manifolds, curved topological spaces that locally, around each point, can be considered similar to a 2D Euclidean space \cite{Chavel2006}. We restrict our interest to continuous and differentiable manifolds where distances and shortest paths can be defined. 

Between any two manifold points, there exists a unique shortest curve in $\mathcal{M}$ that connects these two points. This curve is called a geodesic curve and is equivalent to a straight line on the Euclidean space \cite{Chavel2006}. The length of the geodesic curve defines the distance between the two points in $\mathcal{M}$. Furthermore, for each point $\alpha$ on a manifold $\mathcal{M}$, we can define a plane, passing from that point, which can be understood as a local linearization of the manifold around $\alpha$. This space is called the tangent space of $\mathcal{M}$ at point $\alpha$ and is denoted $\mathcal{T}_{\alpha}\mathcal{M}$. $\mathcal{T}_{\alpha}\mathcal{M}$ has equal dimensionality to the manifold $\mathcal{M}$. 

A tangent vector $u \in \mathcal{T}_{\alpha}\mathcal{M}$ can be uniquely associated to the geodesic curve from point $\alpha$ to point $\beta \in \mathcal{M}$, using the exponential map function $exp : \mathcal{T}_{\alpha}\mathcal{M} \rightarrow \mathcal{M}$, with $exp_{\alpha}(u) = \beta$. The inverse of the exponential map is termed logarithmic map. It accepts two points on the manifold and returns the tangent vector that corresponds to the geodesic curve connecting the two points \cite{DoCarmo1992}, i.e. $\log_{\alpha}(\beta) = u$.

An important inference is that $\log_{\alpha}(\beta)$ is the smallest tangent vector in norm such that $\beta = exp_{\alpha}(u)$ \cite{Pennec1999}. As such, the norm of the logarithmic map provides the length of the geodesic and is used as the distance metric on the manifold:
\begin{equation}\label{eq:riemannian_distance}
d_{\mathcal{M}}(\alpha,\beta) = \| \log_{\alpha}(\beta) \|.
\end{equation}
The gradient of the squared distance function is directly related to the logarithmic map \cite{fletcher2013geodesic}:
\begin{equation}\label{eq:riemannian_gradient}
\nabla_{\alpha} d_{\mathcal{M}}(\alpha,\beta)^2 = -2 \log_{\alpha}(\beta) = -2 u.
\end{equation}

\subsubsection*{Uniform Distributions of Landmarks in Individual Surfaces}

Here we assume that $M$ points have already been positioned on a surface and describe the methodology for distributing these points uniformly on that surface. By maximizing the sample differential entropy $\bar{H}$ w.r.t. $ x_j^1, \ldots , x_j^M$, we in essence manipulate point positions to achieve the required uniformity. The optimization problem can be written as
\begin{equation}\label{argmin2}
\hat{z}_j = \argmin_{z_j \in R^{3 \cdot M}} - \bar{H}_j(x_j^1, \ldots , x_j^M), \  s.t. \  x_j^1, \ldots , x_j^M \in \mathcal{S}_j,
\end{equation}
with $\bar{H}(x_j^1, \ldots , x_j^M) = - \frac{1}{M} \sum_{k=1}^M \ln p(x_j^k)$.

In order to minimize the cost, which is equal to the negative sample entropy, we will employ a gradient descent technique, where points are iteratively moved proportionally to the negative gradient of the cost, until no significant improvement in cost can be achieved. The gradient of $-\bar{H}_j$ w.r.t the landmark point $x_j^k$ is
\begin{equation}\label{der}
\nabla_{x_j^k} ( - \bar{H}_j ) = \nabla_{x_j^k} \left[ - \frac{1}{M} \sum_{l=1}^M \ln p_j(x_j^l) \right] = - \frac{1}{M} \sum_{l=1}^M \frac{\nabla_{x_j^k} p_j(x_j^l) }{p_j(x_j^l)} \approx - \frac{1}{M} \frac{\nabla_{x_j^k} p_j(x_j^k) }{p_j(x_j^k)}.
\end{equation}
The last approximation is based on the assumption that within one iteration of the gradient descent optimization cycle, the probability density at one point is not affected by the rest. As such, $ \nabla_{x_j^k} p_j(x_j^l) = 0$, when $k \neq l$. The assumption is adopted in order to reduce significantly the computational burden. 

To proceed, estimates of $p_j(x_j^l)$ are needed. Kernel Density Estimation is suitable for this purpose, but relies on the calculation of distances between surface points. Polyhedral surfaces can be considered piecewise planar approximations of 2D manifolds. An appropriate formulation of geodesic distances for polyhedral surfaces has been given in the seminal work of Mitchell et al \cite{mitchell1987discrete}. They are computed as overall lengths of piece-wise linear segments on the surface triangles that form straight lines when two adjacent faces are unfolded across their common edge. 

Based on the ability to compute geodesic distances on the surface $S_j$, we propose the following geodesic kernel density estimator for $p_j$:
\begin{equation}
\hat{p}_j(x_j^k) = \frac{1}{M} \sum_{l=1}^M G_{\mathcal{M}} \left( d_{\mathcal{M}}(x_j^k, x_j^l),\sigma_k \right),
\end{equation}
with the isotropic covariance kernel function $G_{\mathcal{M}}: \mathcal{M} \times \mathcal{M} \rightarrow R$ given by
\begin{equation}
G_{\mathcal{M}} \left( d_{\mathcal{M}}(x_j^k, x_j^l),\sigma_k \right) = ( 2\pi \sigma_k )^{-1} e^{-0.5 \sigma_k^{-1} d_{\mathcal{M}}(x_j^k, x_j^l)^2},
\end{equation}
where $\sigma_k$ is a standard deviation parameter.

Having formulated a suitable density estimator, we proceed to solve the optimization problem \eqref{argmin2} using gradient descent. Reminding that points $ x_j^1, \ldots , x_j^M$ are constrained to lie on $S_j$, we need to provide suitable updates for the gradient descent algorithm. Utilizing equation \eqref{eq:riemannian_gradient}, we can write the gradient of the objective function w.r.t. $x_j^k$ as an average vector on the tangent space of the landmark. In particular:

\begin{equation}\label{dc}
\nabla_{x_j^k} ( - \bar{H}_j )  \approx - \frac{1}{M} \frac{\nabla_{x_j^k} \hat{p}_j(x_j^k) }{\hat{p}_j(x_j^k)} = - \frac{1}{\sigma_k^2 } \sum_{l=1}^M  \frac{G_{\mathcal{M}} \left( d_{\mathcal{M}}(x_j^k, x_j^l),\sigma_k \right)}{\sum_{u=1}^M G_{\mathcal{M}} \left( d_{\mathcal{M}}(x_j^k, x_j^u),\sigma_k \right)} \log_{x_j^k}(x_j^l).
\end{equation}

Employing the definition of the exponential map on manifolds, our point updates are given by
\begin{equation}\label{grad1}
x_j^k \leftarrow exp_{x_j^k} \left( - \gamma \nabla_{x_j^k} \bar{H}_j \right),
\end{equation}
where $\gamma$ is the gradient descent's time step parameter. Exponential maps can be computed on polyhedral surfaces using straightest geodesic lines \cite{polthier2006straightest}.

\subsubsection*{Correspondence Optimization}

Having laid out the optimization procedure required to guarantee a uniform distribution of points on the individual surfaces, we now need to solve the correspondence optimization problem, which is written as follows:

\begin{equation}\label{argmin1}
\hat{z}_j = \argmin_{z_j \in R^{3 \cdot M}} \bar{\Gamma}(z_1, \ldots , z_N),  \forall j \in \{1,\ldots,N\}.
\end{equation}

Let again $Z \in R^{3 \cdot M}$ be a random variable in shape space with probability density function $p(Z)$. If the density function is assumed to be a multivariate normal distribution $G(\mu,\Sigma) \in R^{3 \cdot M}$, the differential entropy $H(Z)$ can then be written as $\Gamma(Z) = \frac{1}{2} \ln \{ (2\pi e)^{3M} | \Sigma | \}$  \cite{cates2007shape}.
Using the sample covariance estimator $\hat{\Sigma} = \frac{1}{N-1} \boldsymbol{Y}^T \boldsymbol{Y}$, with $y_j = z_j - \bar{z}$ the centered observations, $\bar{z} = \frac{1}{N} \sum_{j-1}^N z_j$, and $\boldsymbol{Y}, \boldsymbol{Z}$ the $N \times 3M$ matrices of sample vectors $y_j$ and $z_j$ respectively as their rows, the sample entropy $\bar{\Gamma}$ is written as
\begin{equation}
\bar{\Gamma}(z_1, \ldots , z_N) = \ln | \hat{\Sigma} | = \ln | \frac{1}{N-1} \boldsymbol{Y}^T \boldsymbol{Y} |.
\end{equation}

Gradient descent can also be employed here. To simplify computations, the mean estimate $\bar{z}$ is considered fixed during each iteration. With this assumption, the matrix of partial derivatives of $\bar{\Gamma}$ can be written as \cite{cates2007shape}:
\begin{equation}
\frac{\partial \bar{\Gamma}}{\partial \boldsymbol{Z}}  = \frac{\partial \bar{\Gamma}}{\partial \boldsymbol{Y}}  = 2 \boldsymbol{Y} ( \boldsymbol{Y}^T \boldsymbol{Y} )^{-1} \approx 2 \boldsymbol{Y} ( \boldsymbol{Y}^T \boldsymbol{Y} + \alpha \boldsymbol{I})^{-1}.
\end{equation}
A regularization term $\alpha$ is added above since in practice, $\hat{\Sigma}$ will not have full rank. 

In order to accommodate these correspondence updates into our geodesic gradient descent method, we project each update on the plane of the point's corresponding mesh triangle. Let $n_j^k$ be the perpendicular unit normal vector of the $x_j^k$'s current triangle. The tangent vector that maximizes the gradient update on the triangle plane is given by
\begin{equation}\label{grad2}
\nabla_{x_j^k} \bar{\Gamma}  = \frac{\partial \bar{\Gamma}}{\partial x_j^k} - n_j^k \cdot \frac{\partial \bar{\Gamma}}{\partial x_j^k}.
\end{equation}

By adding up the gradients from equations \eqref{grad1} and \eqref{grad2}, we finally acquire the gradient descent updates for the optimization of the overall correspondence cost \ref{combined}:

\begin{equation}
x_j^k \leftarrow exp_{x_j^k} \left( - \gamma \left( \nabla_{x_j^k} \bar{\Gamma} + \nabla_{x_j^k} \bar{H}_j \right) \right)
\end{equation}

\subsubsection*{Initialization and Iterative Landmark Splitting}

The number of landmarks, $M$, to be sampled from each surface is provided as a parameter by the user. The algorithm initially samples randomly one point from each surface and performs correspondence optimization. Following that, the landmark is split into two and the overall optimization process runs again on the new point sets. This procedure is repeated iteratively until the required number of sampled landmarks is reached and their positions are optimized. 

\subsubsection*{Validation of GESSA using the MorphFace dataset}

For this validation, we used the publicly and freely available Morphface dataset of 3D facial surfaces \cite{paysan20093d}. The dataset is distributed freely from the University of Basel for internal, non-commercial research, evaluation or testing purposes, and can be found at \url{http://faces.cs.unibas.ch/bfm/main.php?nav=1-1-1&id=scans}. It comprises of $11$ individual subjects' faces with neutral pose. All the faces were registered to the Basel Face Model (BFM) facial template. The template mesh consisted of $53,490$ landmarks. The registration process identified the locations of these landmarks on the $11$ validation faces. For the purposes of this validation study, $19$ of these landmarks, corresponding to prominent facial points, were manually selected to constitute the groundtruth landmarks (GTLs). Supplementary Figure 8 depicts these $19$ landmarks on the facial surface. The Morphface surfaces were subjected to cropping in order to remove non relevant areas, such as neck and ears. Subsequently, the facial meshes were subsampled, yielding approximately $4,700$ points per face, a number that was similar to that of the facial meshes in our TwinsUK cohort.

Using GESSA, we computed $4,096$ GESSA Generated Landmarks (GGLs) in the validation surfaces. Supplementary Figure 7 (A) depicts these landmarks on three example faces, while Supplementary Figure 9 depicts the average face of the validation dataset. 

The goals of our validation study were threefold. First, we aimed to evaluate how uniform the distribution of the sampled landmark points on the surfaces would be. Second, we aimed to demonstrate that sampling approximately $4000$ facial landmarks produces a dense landmark coverage of facial surfaces. Third, we aimed to show that GESSA is consistently and accurately aligning landmarks to achieve one-to-one correspondence across all surfaces.

First, we focus on the evaluation of the landmarks' surface distributions. Supplementary Figure 11 shows the estimated surface density function of an example validation face. It can be observed that the density is almost uniform everywhere. In addition, examination of Supplementary Figure 7 (A) also shows clearly that the landmarks are uniformly distributed across the complete facial surfaces.

Second, we needed to examine if the chosen number of sampled GGLs is sufficient to densely cover the facial surfaces. A low number of sampled landmarks would lead to extended facial areas having no or very few landmarks. As a consequence, extracted traits would not be able to capture well the morphological information contained in these areas. On the other hand, very dense sampling would incur unnecessary computational costs.

To address this objective, we worked under the premise that, by sampling surfaces densely, we should be able to locate GGLs sufficiently close to the preselected GTLs. We measured the average distance, over all faces in the validation set, from each GTL to its closest GGL. Supplementary Figure 7 (B) shows the GTLs and respective closest GGLs for three example validation faces. Table \ref{table_val1m} reports the $19$ average distances between GTLs and GGLs over the set of $11$ faces. Due to the fact that dimensional units on the Morphface dataset were unknown, we further divided each of the above values by the mean facial width - distance between the two zygomatic landmarks-  in order to compute normalized distances, which are also included in Table \ref{table_val1m}.

The average distances per landmark were always less than $3\%$ of the mean facial width. By further averaging over the $19$ landmarks considered, the overall average distance was $1.77\%$ of the mean facial width. The results indicated that the number of landmarks was sufficient to cover densely the surfaces and that landmarks extended over the complete surfaces.

Third, we concentrated on evaluating the ability of GESSA to place landmarks in equivalent positions consistently across the complete set of faces. The standard deviations of the above GTL-GGL distances can be used as measures of consistent placement, since small values indicate that GTL-GGL distances are similar across all faces and, consequently, landmarks are positioned in equivalent positions across the dataset. The details are reported in Table \ref{table_val1m}. The standard deviations of the GTL-GGL distances for all landmarks were always less than $1\%$ of the mean facial width. By further averaging again over the $19$ landmarks considered, the average standard deviation was $0.42\%$. The results indicated that the automatic landmarking produced by the algorithm was sufficiently accurate to be deployed in our heritability study.

\subsection*{Curvature-based Facial Traits}

The shape around a point on a surface can be characterized using curvature descriptors \cite{koenderink1992surface}. Curvature is a directional property and describes how bent the surface is around each point \cite{roberts2001curvature}. The curvature magnitude of a point in some direction is given by the reciprocal of the radius of the circle that best approximates the slice of surface in that direction \cite{koenderink1992surface}. Normal curvatures are defined on orthogonal planes to a surface point and for each such point, there exists a single normal curvature that has the largest absolute curvature magnitude. This is called the maximum curvature $K_{max}$. The curvature perpendicular to $K_{max}$ is the minimum curvature $K_{min}$. These two surface attributes are collectively called principal curvatures and any normal curvature at a point on a surface can be derived as a combination of  $K_{max}$ and  $K_{min}$ \cite{roberts2001curvature}. Supplementary Fig. 1 
shows a general characterization of shapes based on the signs of $K_{max}$ and $K_{min}$.  Univariate curvature indices derived from these measures have been proposed. Four such measures were used in this study to compute phenotypic traits:

{\bf Mean Curvature: }
\begin{equation}
MC =  \frac{K_{max} + K_{min}}{2} ,
\end{equation}

{\bf Gaussian Curvature: }
\begin{equation}
GC = K_{max} \times K_{min} ,
\end{equation}

{\bf Curvedness: }
\begin{equation}
CU = \sqrt{\frac{K_{max}^2 + K_{min}^2}{2}} ,
\end{equation}

{\bf Shape Index: }
\begin{equation}
SI =  \frac{ \pi }{2} \times \arctan \left( \frac{K_{max} + K_{min}}{K_{max} - K_{min}}  \right) .
\end{equation}

\begin{figure}[ht]
\centerline{\includegraphics[width=5.0in]{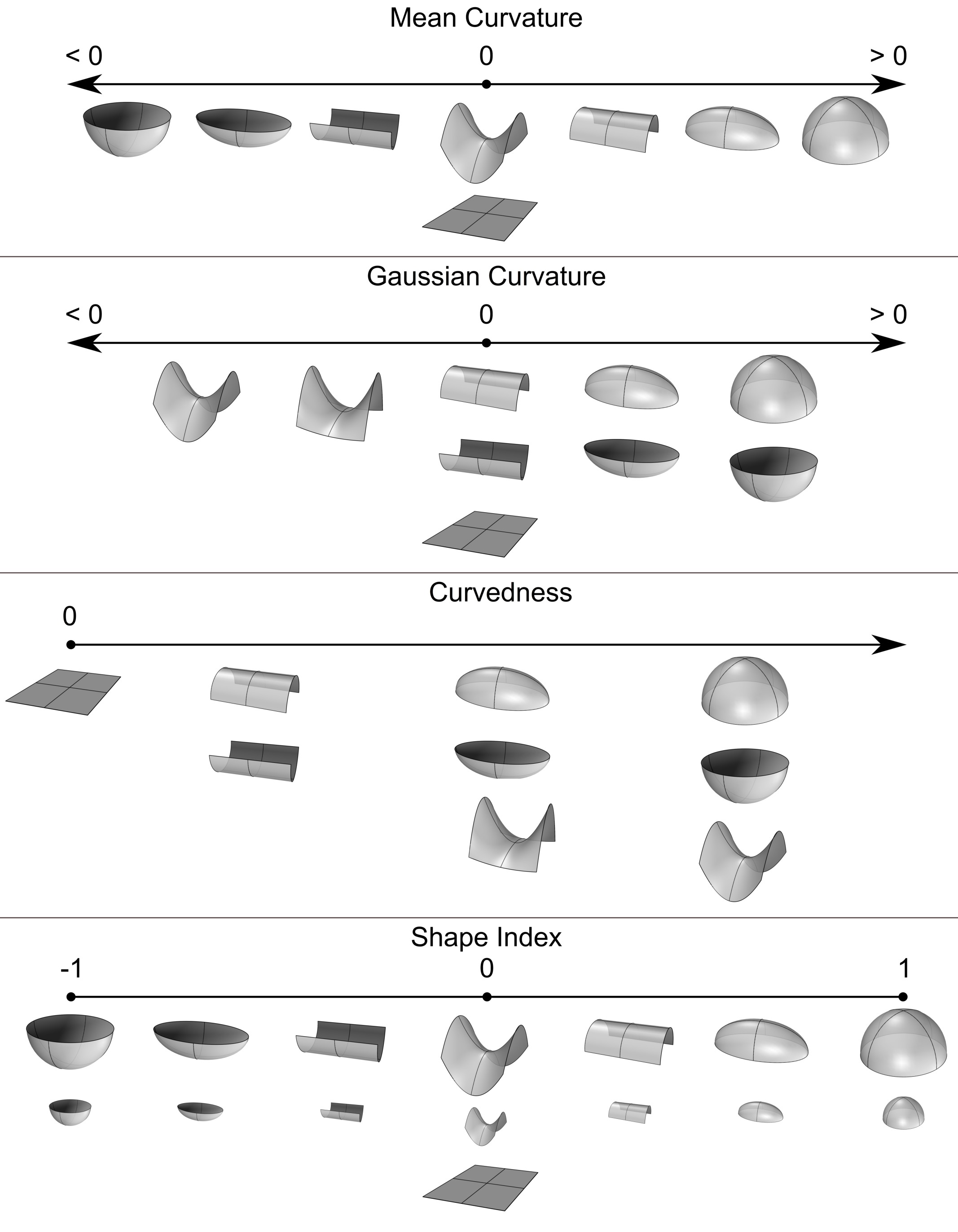}}
\caption{{\bf Topological characteristics of curvature indices.} Each descriptor highlights different attributes of the surface's underlying topology. MC differentiates significantly areas of high and low curvature, as well as convex and concave shapes. GC discriminates well between spherical and saddle-like areas. CU is less representative of a particular morphology and reflects the absolute curvature magnitude in each point, irrespective of its specific shape. Finally, SI is scale-independent and able to differentiate between pure shape characteristics, e.g domes, ridges and saddles, regardless of their high or low CU. }
\label{fig_5}
\end{figure}

In order to explain the differences between the various indices, we first present two intuitive features that can be used so as to describe the shape of a surface patch. The first feature is the general shape morphology and is governed by the various sign combinations of $K_{min}$ and $K_{max}$ , i.e. flat ($K_{max} = K_{max} = 0$ ), vs. convex cylindrical ($K_{max} > 0,  K_{min} = 0$), vs. concave cylindrical ($K_{max} = 0,  K_{min} < 0$), vs saddle structure ($K_{max} > 0,  K_{min} < 0$), vs. convex cylindrical ($K_{max} > 0,  K_{min} > 0$), vs. concave cylindrical ($K_{max} < 0,  K_{min} < 0$). The second feature is curvature magnitude, i.e. how bent the surface is irrespective of shape morphology. Any univariate descriptor of curvature needs to be a compromise, since it cannot include all information provided by these two features \cite{roberts2001curvature, koenderink1992surface}. Individually, each principal curvature does not provide a useful interpretation of local surface shape \cite{koenderink1992surface}, as can be seen in Supplementary Fig. 1. 
In contrast, the four curvature indices yield more meaningful quantitative shape measures by grouping together or differentiating particular classes of basic shape structures. An illustration of the various indices' characteristics, including their domains and how shapes are differentiated are included in Fig. \ref{fig_5}. The MC index provides a balanced measure between shape morphology and curvature magnitude. It is strongly affected by directional shape differences (convex vs. concave shapes) but is also sensitive to curvature magnitude. GC distinguishes primarily between shape morphologies of the same or opposite principal curvature signs. Finally, CU and SI indices are the most accurate quantitative measures of curvature magnitude and shape morphology respectively.

The calculation of local curvature on the 3D meshes was performed using a finite-differences algorithm \cite{rusinkiewicz2004estimating}. Normal vectors perpendicular to the surface were computed in each landmark. Gradients across the surface were then approximated using finite normal differences of neighbor points. Principal curvatures were found through an eigenvalue decomposition of the normal gradients and curvature index values calculated from the respective formulas.

\subsection*{Distance-based Facial Traits}

Two different types of distance-based phenotypes were considered in this study. Traits derived from Euclidean distances between landmark pairs (EDTs), which represent the main type of examined phenotypes in the literature, and traits derived from Geodesic distances between landmark pairs (GDTs). An illustration of the difference between the two types of distances can be seen in Supplementary Fig. 6. 
Geodesic computations were performed using an implementation of the exact discrete geodesic algorithm \cite{mitchell1987discrete, surazhsky2005fast}.

\subsection*{Heritability Estimation}

The heritability analyses were performed using Structural Equation Modeling (SEM) \cite{rijsdijk2002analytic}. SEM evaluates which combination of additive (A) genetic, common (C) environmental and unique (E) environmental variance components can best explain the observed phenotypic variance and covariance of MZ and DZ twin data. The importance of individual variance components is assessed by dropping parameters sequentially from the set of nested models ACE$\rightarrow$AE$\rightarrow$E. In choosing between models, variance components are excluded from the selection process if there is no significant deterioration in model fit after the component is dropped, as assessed by the Akaike Information Criterion (AIC) \cite{akaike1974new}. The E component represents random error and must be retained in all models \cite{rijsdijk2002analytic}. Heritability estimates for the AE models are calculated as $\frac{a^2}{a^2 + e^2}$, where $a$ and $e$ are the path coefficients of the A and E variance components in the SEM model. A detailed description of SEM can be found in the Supplementary Text. In this work, ACE, AE and E structural equation models were fitted using the OpenMx software \cite{boker2011openmx}. Regarding the univariate heritability studies, the log files from OpenMx for MC, GC, CU and SI traits respectively are provided in: Supplementary Files 5-8 for the ACE models, Supplementary Files 9-12 for the AE models and Supplementary Files 13-16 for the E models. Regarding the multivariate heritability studies, the log files (ACE, AE and E models) from OpenMx for MC, GC, CU and SI traits respectively are provided in Supplementary Files 17-20. The log files include detailed fit statistics, as well as the estimated path coefficients for the latent factors, along with their standard errors. Furthermore, age was included in the SEM models as a covariate. In contrast, for the multivariate study, the age was regressed from the data before the application of SEM. While the outcome is identical for both approaches, in the second case, no age variable is shown in the models, and ACE, AE and E logs are all included in single files.

\section*{Acknowledgements}

The study was funded by the Wellcome Trust. It was supported by the National Institute for Health Research (NIHR) BioResource, Clinical Research Facility and Biomedical Research Center based at Guy's and St Thomas' NHS Foundation Trust in partnership with King's College London, as well as the International Visible Trait Genetics (VisiGen) Consortium (Timothy D. Spector, Department of Twin Research and Genetic Epidemiology, King’s College London, United Kingdom. Pirro G. Hysi, Department of Twin Research and Genetic Epidemiology, King’s College London, United Kingdom. Fan Liu, Department of Genetic Identification, Erasmus MC University Medical Center Rotterdam, The Netherlands. Lavinia Paternoster, MRC Integrative Epidemiology Unit, School of Social and Community Medicine, University of Bristol, United Kingdom. David M. Evans, MRC Integrative Epidemiology Unit, School of Social and Community Medicine, University of Bristol, United Kingdom. Stefan Boehringer, Department of Medical Statistics and Bioinformatics, Leiden University Medical Center, The Netherlands. David L. Duffy, QIMR Berghofer Medical Research Institute, Brisbane, Australia. Nicholas G. Martin, QIMR Berghofer Medical Research Institute, Brisbane, Australia. Manfred Kayser, Department of Genetic Identification, Erasmus MC University Medical Center Rotterdam, The Netherlands). Visigen is sponsored by Identitas Inc. TwinsUK is funded by the Wellcome Trust, Medical Research Council, European Union, the National Institute for Health Research (NIHR)-funded BioResource, Clinical Research Facility and Biomedical Research Centre based at Guy’s and St Thomas’ NHS Foundation Trust in partnership with King’s College London. P.H. is the recipient of a Fight for Sight Early Career Investigator Award and T.S. is an NIHR Senior Investigator.

\section*{Author contributions}

D.T., P.H., G.M. and T.S. conceived the study and analyzed the results. D.T. conducted the experiments. All authors reviewed the manuscript. 

\section*{Additional information}

\subsection*{Competing financial interests} 

The authors declare no competing financial interests.

\FloatBarrier

\begin{table}[ht]
\centering
\begin{tabular}{|c|c|c|c|}
\hline
\multicolumn{2}{|c|}{Landmarks} & \multicolumn{2}{|c|}{Distance Phenotypes} \\
\hline
Abbreviations & Description &  Abbreviations & Description \\
\hline
g & Glabella & sn - gn & Height of lower face \\
n & Nasion  & g - sn & Height of middle face \\
prn & Pronasale & n - sn & Nose Height \\
sn & Subnasale & sn - prn & Nasal protrusion \\
gn & Gnathion & alL - alR & Nose width \\
alL & Alare (L) & exL - exR & Intercanthal width \\
alR & Alare (R) & enL - enR & Biocular width \\
exL & Exocanthion (L)  & zyL - zyR & Zygomatic width\\
exR & Exocanthion (R) & goL - goR & Mandible width\\
enL & Endocanthion (L) & chL - chR & Mouth width\\
enR & Endocantion (R) & & \\
zyL & Zygion (L) & & \\
zyR & Zygion (R) & & \\
chL & Cheilion (L) & & \\
chR & Cheilion (R) & & \\
goL & Gonion (L) & & \\
goR  & Gonion (R) & & \\
\hline
\end{tabular}
\caption{\label{table_1} Abbreviations and related descriptions of selected landmarks and distance-based traits.}
\end{table}

\begin{table}[ht]
\centering
\begin{tabular}{|c|c|c|c|c|}
\hline
Phenotypic Traits & \multicolumn{2}{|c|}{EDT} & \multicolumn{2}{|c|}{GDT} \\
\cline{2-5}
& $h^2$ & Goodness-Of-Fit test $p$-value &  $h^2$ & Goodness-Of-Fit test $p$-value \\
\hline
sn - gn & 0.634  & 0.019  & {\bf 0.692} & {\bf 0.126} \\
g - sn & 0.75  & 0.00003  & 0.708  & 0.0004    \\
n - sn & 0.749 & 0.00001 & 0.699 & 0.0003  \\
sn - prn & {\bf 0.545}  & {\bf 0.276} & {\bf 0.505} & {\bf 0.731}  \\
alL - alR & {\bf 0.718}  & {\bf 0.219} & 0.697 & 0.0002  \\
exL - exR & 0.706  & 0.005 & 0.651  & 0.0002   \\
enL - enR & 0.707  & 0.02  & {\bf 0.789}  &  {\bf 0.063}  \\
zyL - zyR & {\bf 0.734}  & {\bf 0.303}  & 0.665 &  0.001 \\
goL - goR & {\bf 0.677}  & {\bf 0.537}  & {\bf 0.573}  & {\bf 0.468}  \\
chL - chR & {\bf 0.62}  & {\bf 0.105}  & {\bf 0.586}  & {\bf 0.449}  \\
\hline
\end{tabular}
\caption{\label{table_2} Heritability estimates $h^2$ for $10$ traits based on Euclidean Distances (EDTs) and $10$ traits based on Geodesic Distances (GDTs). Models' Goodness-Of-Fit test $p$-values greater that $0.05$ correspond to good fits of the observed data.}
\end{table}

\begin{table}[ht]
\centering
\begin{tabular}{|c|c|c|c|c|}
\hline
Landmarks & Average Distance & Normalized Average Distance & Std & Normalized Std \\
\hline
Zygion (R) &  1399.047 & 0.01077 & 422.24 & 0.00325\\
Labiale Inferius & 1791.735 & 0.01379 & 472.386 & 0.00363\\
Sublabiale &  2162.848 & 0.01665 & 330.163 &  0.00254\\
Gnathion  & 2506.586 &  0.01929 & 480.526 & 0.00369\\
Glabella & 1677.904 & 0.01291 & 555.001 & 0.00427\\
Labiale Superius &  2545 & 0.01959 & 594.255 & 0.00457\\
Cheilion (R)  & 2274 & 0.0175 & 542.486 & 0.00417\\
Nasion  & 2896.955 & 0.0223 & 808.766 & 0.00622\\
Pronasale & 1625.78 & 0.0125 & 557.833 & 0.00429\\
Subnasale  & 2389.583 & 0.01839 & 777.568 & 0.00598\\
Alare (R)  & 3694.01 & 0.02844 & 678.1579 &  0.00522\\
Endocantion (R) & 1526.617 &  0.01175 & 331.483 & 0.00255\\
Exocanthion (R) &  2152.691 & 0.01657 & 487.591 & 0.00375\\
Progonion & 3008.975 & 0.02316 & 574.951 & 0.00442\\
Zygion (L)  & 3142.033 & 0.02419 & 522.777 & 0.00402\\
Cheilion (L)  & 2571.254 & 0.01979 & 583.14 & 0.00448 \\
Alare (L)  & 2093.014 & 0.01611  & 514.262 & 0.00395\\
Endocanthion (L) & 2420.324 & 0.01863 & 367.685 & 0.00283\\
Exocanthion (L) & 1771.4 & 0.01363 & 736.372 & 0.00566\\

\hline
\end{tabular}
\caption{\label{table_val1m} Average distances and standard deviations (Std) between the $19$ groundtruth landmarks and their respective nearest GESSA landmark points. Statistics were computed on the set of $11$ validation faces. Normalized measurements are acquired by division with the mean facial width.}
\end{table}

\FloatBarrier

\newpage

\appendix

\renewcommand\thefigure{\thesection.\arabic{figure}}    
\setcounter{figure}{0}
\renewcommand\thetable{\thesection.\arabic{table}}    
\setcounter{table}{0}

\section{Supplementary Information}

\subsection{Supplementary Figures}

\begin{figure}[ht]
\centerline{\includegraphics[width=4.0in]{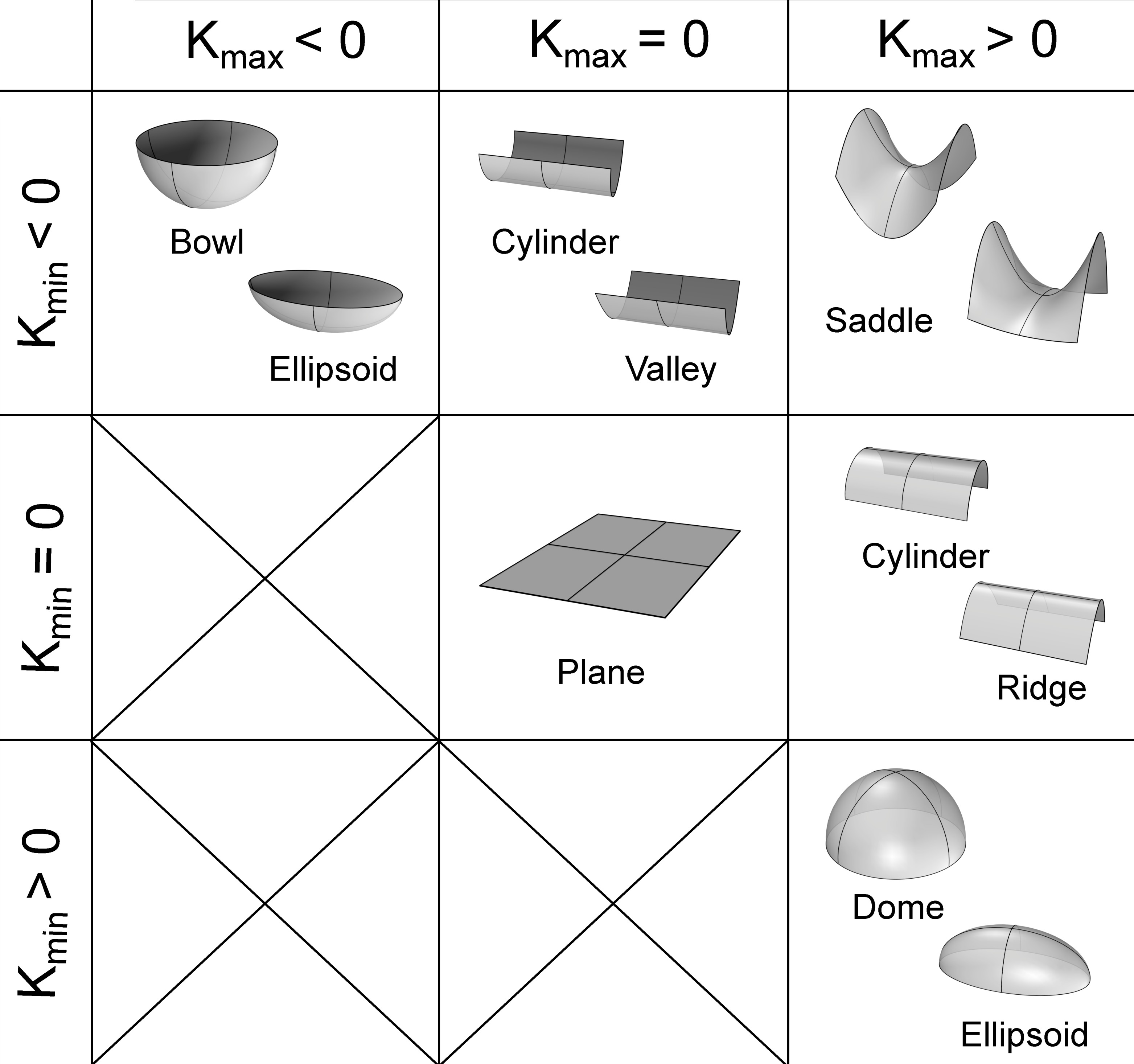}}
\caption{{\bf Principal curvatures and shape characterization.} General classification of shapes based on the signs of the two principal curvatures. While principal curvatures include all information about the curvature at a point, both numbers are needed in order to get meaningful shape categorizations.
}
\label{fig_SB}
\end{figure}

\begin{figure}[ht]
\centerline{\includegraphics[width=4.5in]{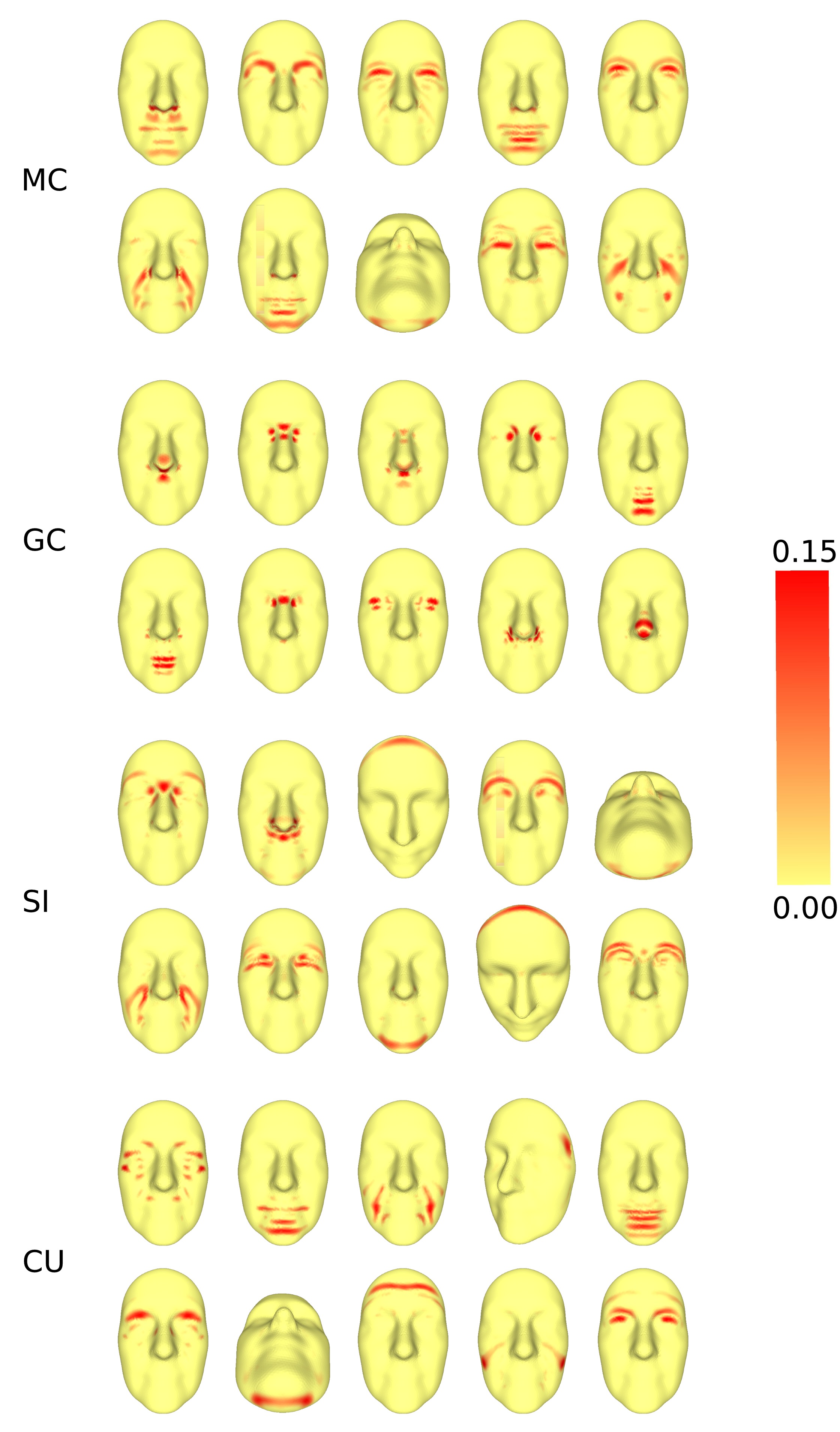}}
\caption{ {\bf Eigenfaces associated to the largest principal components for each curvature index.} For each curvature index, the faces are arranged in decreasing order, from left to right. 
}
\label{fig_SC}
\end{figure}

\begin{figure}[ht]
\centerline{\includegraphics[width=7.0in]{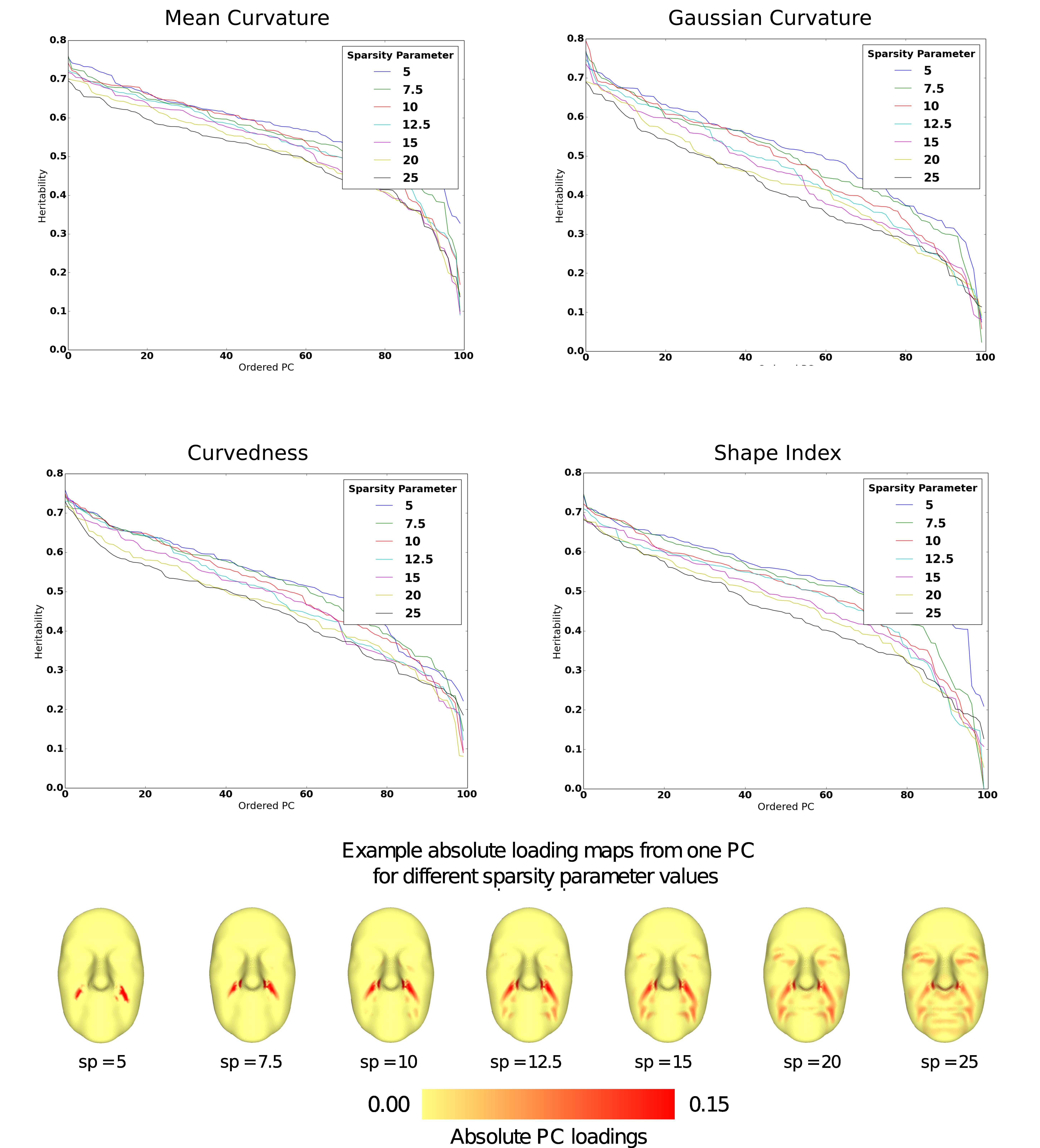}}
\caption{{\bf Effect of sparsity parameter on composite curvature trait heritability analysis. } Different sparsity parameters were tested in order to assess the parameter's contribution on the heritability study. Each plot shows the sorted heritability values for the first $100$ sPCs per curvature descriptor and for $7$ different sparsity parameters. Heritability estimates showed similar behavior irrespective of how sparse the components were. The facial maps depict the absolute loadings of an example composite trait (Mean curvature descriptor) for different sparsity parameter values and their computed heritability values.
}
\label{fig_SD}
\end{figure}

\begin{figure}[ht]
\centerline{\includegraphics[width=4in]{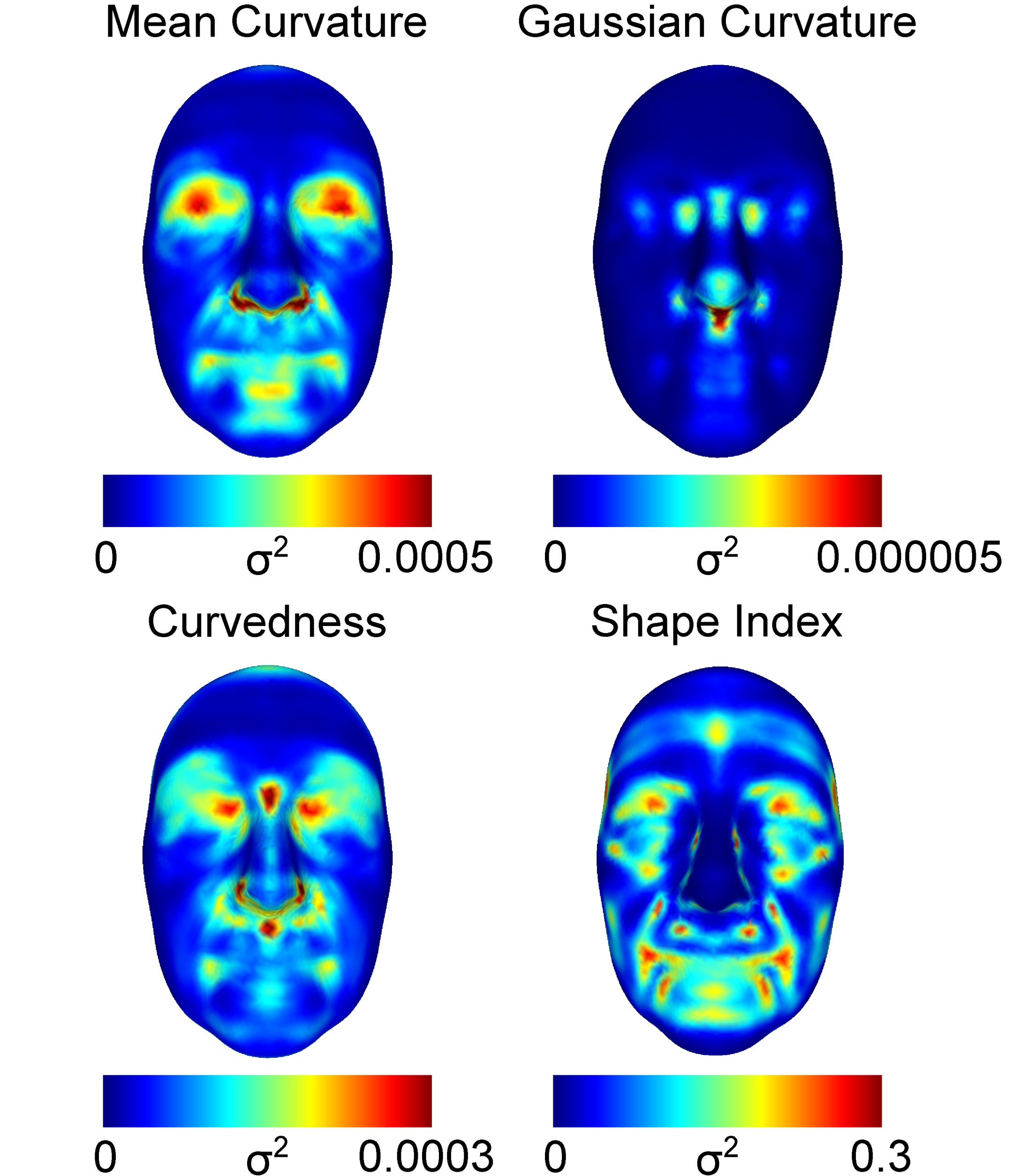}}
\caption{{\bf Curvature variance Maps. } The maps illustrate the variance of curvature values on all landmarks, computed from the full dataset of $952$ TwinsUK subjects.
}
\label{fig_SE}
\end{figure}

\begin{figure}[ht]
\centerline{\includegraphics[width=6.5in]{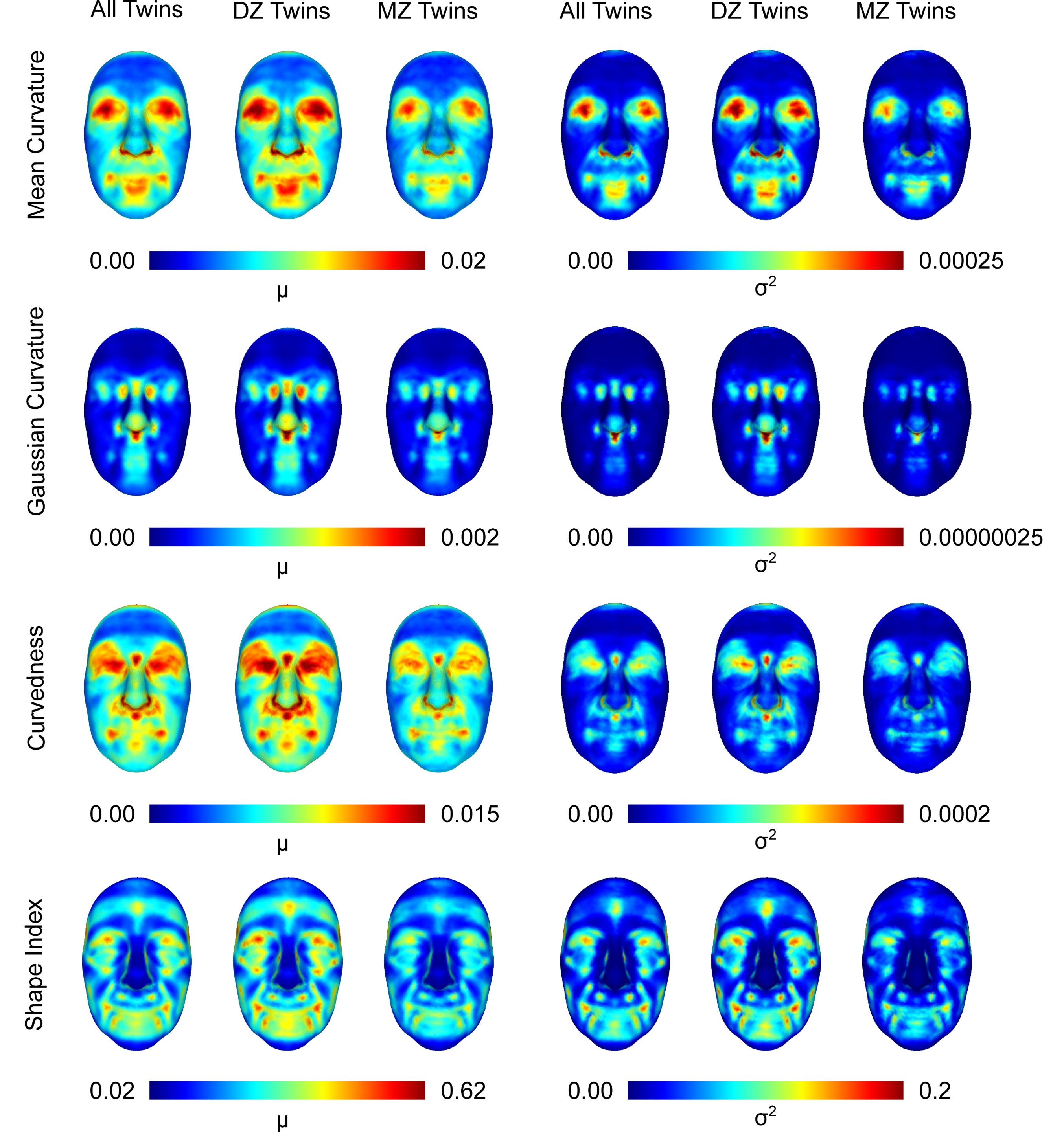}}
\caption{{\bf Mean and variance maps of absolute curvature differences between twins. } For each pair of twins, the absolute difference in curvature values was computed on all facial points. The maps show the mean and variance of the differences for all twins, as well as for MZ and DZ subsets.
}
\label{fig_SF}
\end{figure}

\begin{figure}[ht]
\centerline{\includegraphics[width=3.0in]{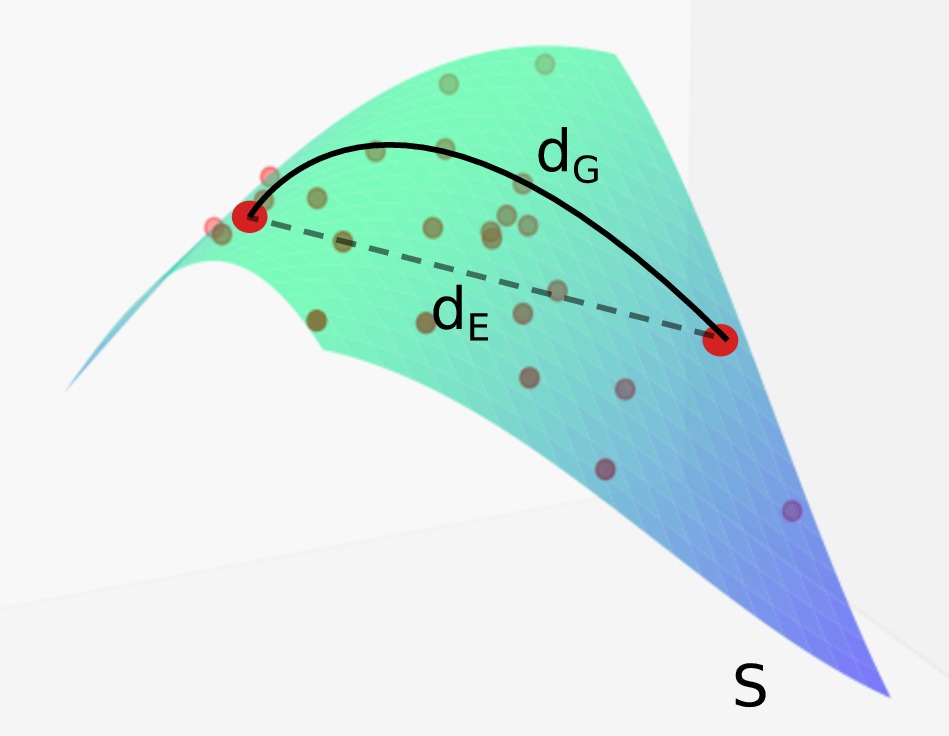}}
\caption{{\bf Illustration example of Geodesic and Euclidean distance metrics. } The distance between two points that lie on a surface $\mathcal{S} \in \mathbb{R}^3$ may be either the length $d_E$ of the straight path between the two points, or the length $d_G$ of the shorted curved path between the same points, under the constraint that movement is only allowed on the surface $\mathcal{S}$.  
}
\label{fig_SG}
\end{figure}

\begin{figure}[ht]
\centerline{\includegraphics[width=4.5in]{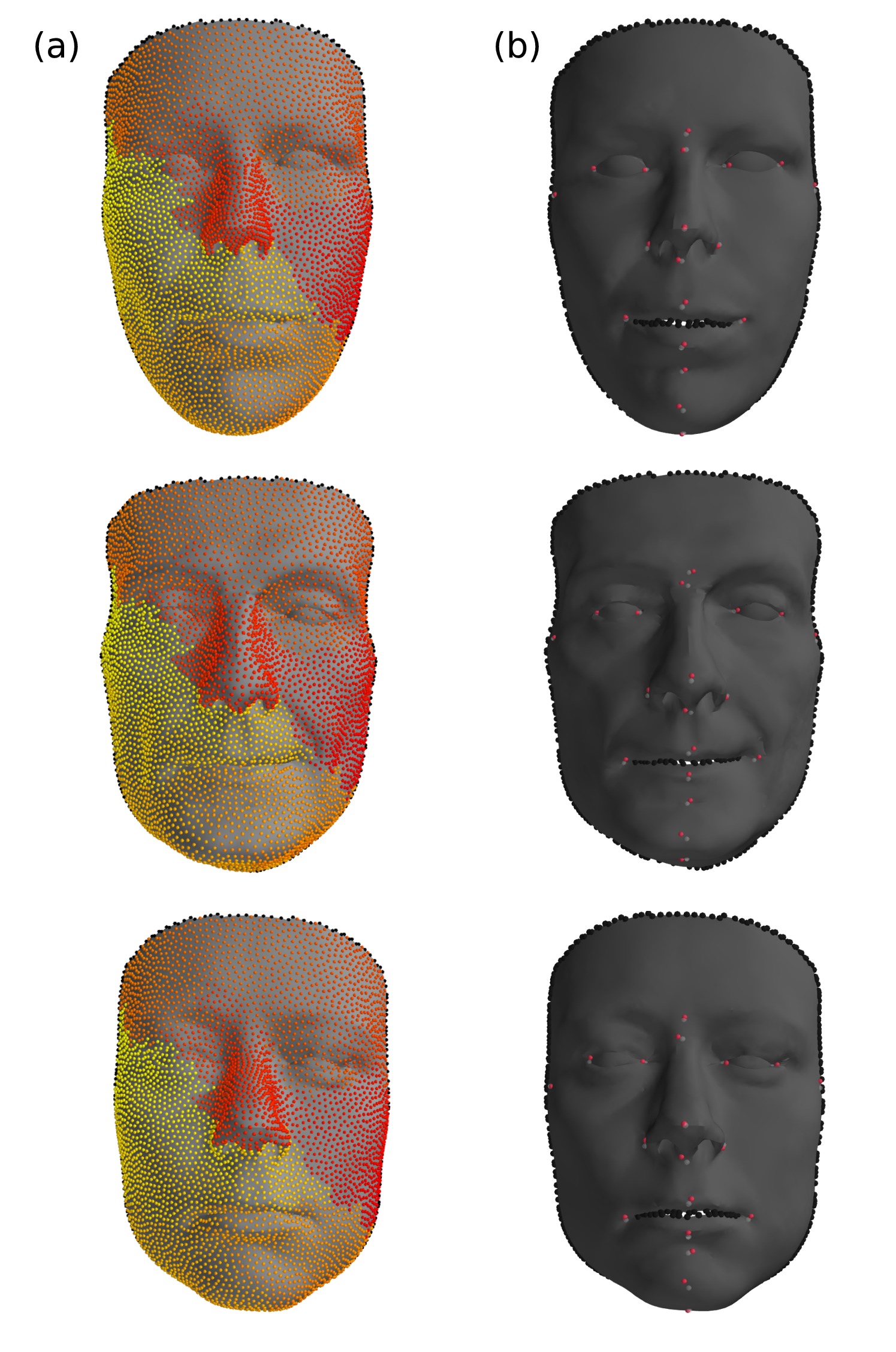}}
\caption{
{\bf Extracted landmark sets on validation faces using GESSA.} (A) $4,096$ corresponding facial points were computed using our dense automated landmarking method. Results for three validation faces are shown here. Corresponding points are colored consistently among the different faces. Preselected Groundtruth landmarks (GTLs) are shown in white.
 (B) Red points represent the nearest GESSA generated landmarks (GESLs) to the GTLs for the same example faces. Dense annotation allowed selection of nearest landmarks with distances from GTLs less than $3\%$ of the mean facial width. Importantly, the algorithm was able to consistently place landmarks across different faces.}
\label{fig_val2mm}
\end{figure}

\begin{figure}[ht]
\centerline{\includegraphics[width=3.0in]{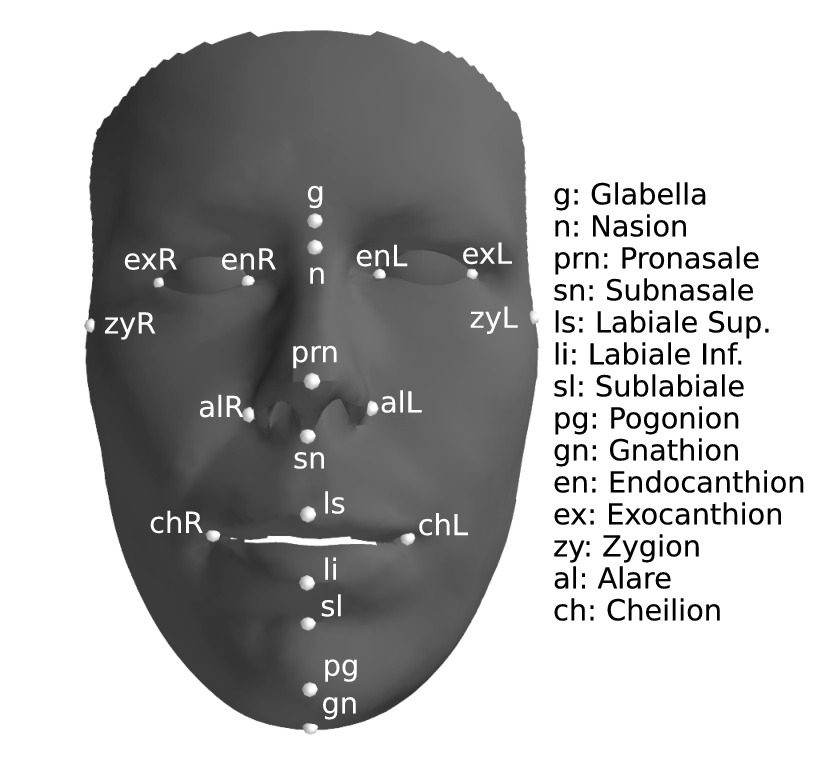}}
\caption{{\bf Morphface Validation Dataset. Example 3D facial surface with groundtruth landmarks.} The $19$ groundtruth landmark positions were used during validation of our GESSA landmarking methodology. 
}
\label{fig_val1m}
\end{figure}

\begin{figure}[ht]
\centerline{\includegraphics[width=1.7in]{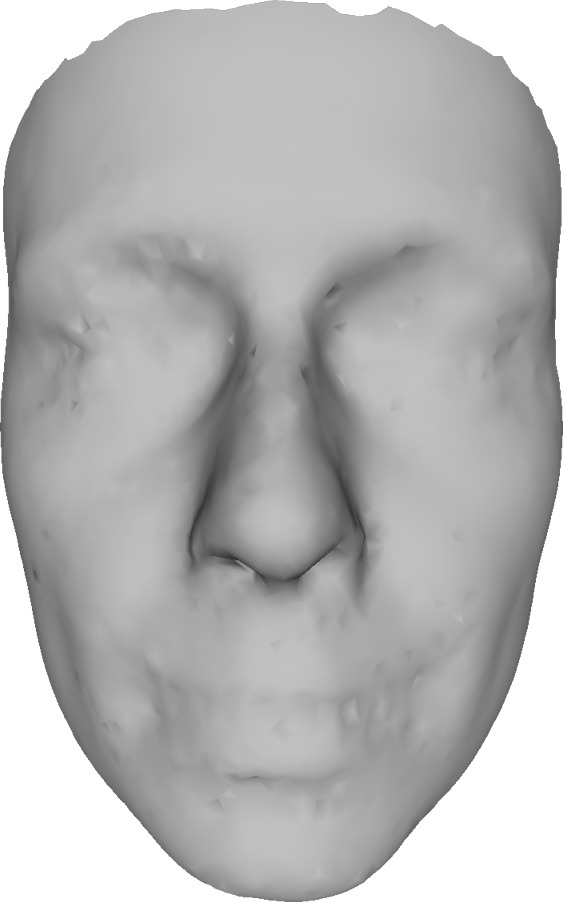}}
\caption{
{\bf Morphface Average Facial Surface.} The average surface was constructed by averaging landmark coordinates of the $11$ validation faces from the Morphface dataset. 
}
\label{fig_val3m}
\end{figure}

\begin{figure}[ht]
\centerline{\includegraphics[width=6.0in]{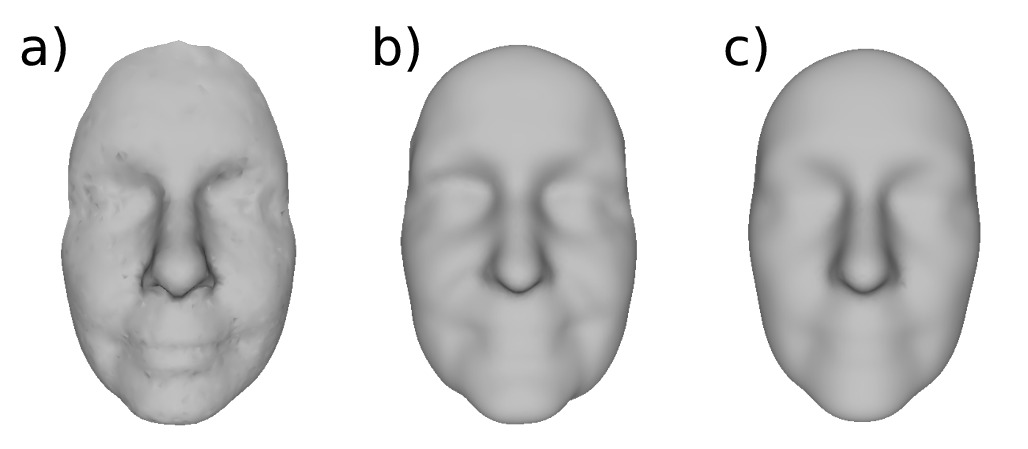}}
\caption{
    {\bf TwinsUK Average Facial Surfaces.} Three average facial surfaces from the TwinsUK dataset using (a) $10$ randomly selected individuals, (b) $200$ randomly selected individuals and (c) the complete dataset. Increasing the number of faces results in a smoother average facial surface.
}
\label{fig_val4m}
\end{figure}

\begin{figure}[ht]
\centerline{\includegraphics[width=2.3in]{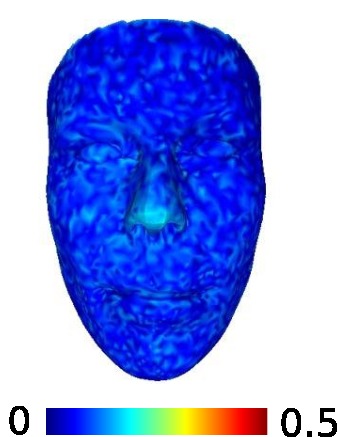}}
\caption{
    {\bf Estimated Probability Density Function on a Validation Face.} In GESSA, landmarks are considered samples of a random variable defined on the facial surface. By maximizing the sample differential entropy (see Equation (7) in the article), the density function becomes almost constant everywhere. The effect of this optimization is that landmarks are effectively drawn from the uniform distribution.
}
\label{fig_val5m}
\end{figure}

\FloatBarrier
\subsection{Supplementary Tables}

\begin{table}[!ht]
\centering
\begin{tabular}{|l|l|l|l|l|l|l|l|l|l|}
\hline
 & & \multicolumn{4}{|l|}{Landmark Curvature Phenotypes} & \multicolumn{4}{|l|}{Composite Curvature Phenotypes}\\ \cline{3-10}
 & & MC & GC & CU & SI & MC & GC & CU & SI \\ \hline

\multirow{ 5}{*}{ACE} 
& $df$ & 5 & 5 & 5 & 5 & 6 & 6 & 6 & 6 \\ \cline{2-10}
& $\chi^2$ & -0.376 & 12.348 & -2.385 & -0.078 & 9.552 & 10.263 & 9.25 & 10.927 \\ \cline{2-10}
& $p$ & 0.573 & 0.442 & 0.576 & 0.603 & 0.29 & 0.273 & 0.29 & 0.251 \\ \cline{2-10}
& $-2\log L$ & 2169.791 & -4186.128 & 1919.943 & 6730.955 & 4850.608 & 4859.602 & 4862.932 & 4868.236 \\ \cline{2-10}
& $AIC$ & 2179.791 & -4176.128 & 1929.943 & 6740.955 & 4858.608 & 4867.602 & 4870.932 & 4876.236 \\ \hline
\multirow{ 5}{*}{AE} 
& $df$ & 6 & 6 & 6 & 6 & 7 & 7 & 7 & 7 \\ \cline{2-10}
& $\chi^2$ & 0.15 & 12.954 & -1.99 & 0.274 & 10.029 & 11.062 & 9.681 & 11.254 \\ \cline{2-10}
& $p$ & 0.592 & 0.459 & 0.599 & 0.63 & 0.326 & 0.303 & 0.33 & 0.292  \\ \cline{2-10}
& $-2\log L$ & 2170.317 & -4185.521 & 1920.339 & 6731.308 & 4851.084 & 4860.402 & 4863.362 & 4868.563 \\ \cline{2-10}
& $AIC$ & 2178.317 & -4177.521 & 1928.338 & 6739.308 & 4857.084 & 4866.402 & 4869.362 & 4874.563 \\ \hline
\multirow{ 5}{*}{E} 
& $df$ & 7 & 7 & 7 & 7 & 8 & 8 & 8 & 8 \\ \cline{2-10}
& $\chi^2$ & 70.223 & 66.951 & 56.491 & 41.964 & 99.493 & 91.209 & 86.867 & 83.24 \\ \cline{2-10}
& $p$ & 0.0195 & 0.03 & 0.073 & 0.111 & 0.002 & 0.002 & 0.002 &  0.01 \\ \cline{2-10}
& $-2\log L$ & 2240.391 & -4131.524 & 1978.82 & 6772.997 & 4940.549 & 4940.549 & 4940.549 & 4940.549 \\ \cline{2-10}
& $AIC$ & 2246.391 & -4125.524 & 1984.82 & 6778.997 & 4944.549 & 4944.549 & 4944.549 & 4944.549 \\ \hline

\end{tabular}
\caption{Model-Fitting Average Statistics for the Curvature-based Heritability Analyses}
\label{tableS1}
\end{table}

\begin{table}[!ht]
\centering

\begin{tabular}{|l|l l|l l|l l|l l|}
\hline
& \multicolumn{2}{|l|}{MC} & \multicolumn{2}{|l|}{GC} & \multicolumn{2}{|l|}{CU} & \multicolumn{2}{|l|}{SI} \\ \hline

\multirow{10}{*}{\rotatebox[origin=c]{90}{Pearson's $r$ } } & \multirow{2}{*}{$6th$ sPC} &  MZ 0.720 & \multirow{2}{*}{$3rd$ sPC} &  MZ 0.709 & \multirow{2}{*}{$17th$ sPC} &  MZ 0.750 & \multirow{2}{*}{$22th$ sPC} &  MZ 0.703 \\
&  &  DZ 0.441 & &  DZ 0.402 &  &  DZ 0.411 & &  DZ 0.407 \\ \cline{2-9}
& \multirow{2}{*}{$43th$ sPC} &  MZ 0.714 & \multirow{2}{*}{$9th$ sPC} &  MZ 0.745 & \multirow{2}{*}{$51th$ sPC} &  MZ 0.737 & \multirow{2}{*}{$27th$ sPC} &  MZ 0.704 \\
&  &  DZ 0.268 & &  DZ 0.450 &  &  DZ 0.351 & &  DZ 0.250 \\ \cline{2-9}
& \multirow{2}{*}{$58th$ sPC} &  MZ 0.704 & \multirow{2}{*}{$67th$ sPC} &  MZ 0.737 & \multirow{2}{*}{$11th$ sPC} &  MZ 0.727 & \multirow{2}{*}{$14th$ sPC} &  MZ 0.692 \\
&  &  DZ 0.363 & &  DZ 0.464 &  &  DZ 0.283 & &  DZ 0.365 \\ \cline{2-9}
& \multirow{2}{*}{$11th$ sPC} &  MZ 0.681 & \multirow{2}{*}{$15th$ sPC} &  MZ 0.694 & \multirow{2}{*}{$8th$ sPC} &  MZ 0.686 & \multirow{2}{*}{$2nd$ sPC} &  MZ 0.644 \\
&  &  DZ 0.343 & &  DZ 0.382 &  &  DZ 0.428 & &  DZ 0.445 \\ \cline{2-9}
& \multirow{2}{*}{$60th$ sPC} &  MZ 0.709 & \multirow{2}{*}{$4th$ sPC} &  MZ 0.695 & \multirow{2}{*}{$62nd$ sPC} &  MZ 0.697 & \multirow{2}{*}{$1st$ sPC} &  MZ 0.666 \\
&  &  DZ 0.343 & &  DZ 0.246 &  &  DZ 0.412 & &  DZ 0.338 \\ \hline

\end{tabular}
\caption{Top Heritable Composite Curvature Traits - Phenotypic Correlations for MZ and DZ subsets}
\label{tableS2}
\end{table}

\begin{table}[!ht]
\centering
\begin{tabular}{|l|l|l|l|}
\hline
 & & \multicolumn{1}{|l|}{Linear Distance Phenotypes} & \multicolumn{1}{|l|}{Geodesic Distance Phenotypes}\\ \hline

\multirow{ 5}{*}{ACE} 
& $df$ & 6 & 6 \\ \cline{2-4}
& $\chi^2$ & 13.458 & 15.757 \\ \cline{2-4}
& $p$ & 0.1941 & 0.1678 \\ \cline{2-4}
& $-2\log L$ & 4775.921 & 4834.57  \\ \cline{2-4}
& $AIC$ & 2879.921  & 2938.569  \\ \hline
\multirow{ 5}{*}{AE} 
& $df$ & 7 & 7 \\ \cline{2-4}
& $\chi^2$ & 16.4  & 17.444 \\ \cline{2-4}
& $p$ & 0.1489  & 0.184  \\ \cline{2-4}
& $-2\log L$ & 4778.863 & 4836.256  \\ \cline{2-4}
& $AIC$ & 2878.862  & 2938.256  \\ \hline
\multirow{ 5}{*}{E} 
& $df$ & 8 & 8 \\ \cline{2-4}
& $\chi^2$ & 179.8 & 163.724 \\ \cline{2-4}
& $p$ & 0 & 0 \\ \cline{2-4}
& $-2\log L$ & 4942.263 & 4982.537  \\ \cline{2-4}
& $AIC$ & 3042.263 & 3082.536  \\ \hline

\end{tabular}
\caption{Model-Fitting Average Statistics for the Distance-based Heritability Analysis}
\label{tableS0}
\end{table}

\begin{table}[!ht]
\centering
\begin{adjustwidth}{-0.25in}{0in} 

\begin{tabular}{|l|l|l|l|l| l l l|}
\hline

Phenotype & Ref. & Study & Sample Size & Ethnicity & Related Maps  &  &   \\ \hline \hline

NW & \cite{kim2013heritabilities} & FBH & 229 & Korean & MC Heritability Map &  &  \\ \hline
IED & \cite{kim2013heritabilities} & FBH & 229 & Korean & 67th GC sPC & 4th GC sPC & 11th CU sPC \\ \hline
NP & \cite{weinberg2013heritability} & TH & 42 & American & 67th GC sPC & 4th GC sPC & 11th CU sPC \\ \hline
FW & \cite{baydacs2007heritability} & TH & 138 & Asian & MC Heritability Map & SI Heritability Map &  \\ \hline
FW & \cite{sherwood2008quantitative} & FBH & 607 & American & MC Heritability Map & SI Heritability Map &  \\ \hline
FW & \cite{ermakov2005quantitative} & FBH & 1406 & European & MC Heritability Map & SI Heritability Map &   \\ \hline
FW & \cite{karmakar2007genetic} & FBH & 373 & Indian & MC Heritability Map & SI Heritability Map &   \\ \hline
HC & \cite{ermakov2010family} & FBH &  1042 & European & CU Heritability Map & & \\ \hline
HC & \cite{karmakar2007genetic} & FBH & 373 & Indian & CU Heritability Map & & \\ \hline
MRL & \cite{johannsdottir2005heritability} & FBH & 363 & European & CU Heritability Map & 60th MC sPC  & 14th SI sPC \\ \hline
MBL & \cite{johannsdottir2005heritability} & FBH & 363 & European & CU Heritability Map & & \\ \hline
MR-MB & \cite{johannsdottir2005heritability} & FBH & 363 & European & CU Heritability Map & & \\ \hline
MR-MB & \cite{carels2001quantitative} & TH & 77 & European & CU Heritability Map & & \\ \hline
CW & \cite{kim2013heritabilities} & FBH & 229 & Korean & CU Heritability Map & 15th GC sPC & 8th CU sPC \\ \hline

\multicolumn{8}{|l|}{ \textbf{NW.} Node Width \textbf{IED.} Inner Eye Corner Distance \textbf{NP.} Nasion Protrusion \textbf{FW.} Face Width} \\ 
\multicolumn{8}{|l|}{ \textbf{HC.} Head Circumference \textbf{MRL.} Mandible Ramus Length \textbf{MBL.} Mandible Body Length} \\
\multicolumn{8}{|l|}{ \textbf{MR-MB.} Mandible Ramus - Mandible Body Angle \textbf{CW.} Chin Width} \\ \hline
\multicolumn{8}{|l|}{ \textbf{FBH.} Family-Based Heritability \textbf{TH.} Twin Heritability} \\ \hline

\end{tabular}
\caption{Comparison between previously reported heritable phenotypes, heritability maps and composite curvature traits}
\label{tableS3}
\end{adjustwidth}
\begingroup
\endgroup

\end{table}

\FloatBarrier
\subsection{Supplementary Text}\label{text_S1}

\subsubsection{Previous Work on Dense Landmarking}

Accurate correspondence of landmark points across surfaces is paramount for shape analysis. Until recently, the primary approach to landmark annotation was to manually localize a small number of landmarks on every data object. In the last years, a number of semi-automated and automated landmarking algorithms have appeared in the literature, which attempt to address these issues. In the following we concentrate specifically to methodologies that compute dense landmark correspondences between surfaces, meaning that thousands of landmark points are annotated in each object. This large number of landmarks allows morphological variability to be captured and quantified in a much more granular level, compared to sparse annotations.

Dense landmarking methods can be classified into one of two categories, based on whether annotation is performed in a pairwise fashion - each surface is registered to a common template, with ensemble correspondence extracted implicitly from all one-to-one surface annotations -  or a groupwise fashion - all surfaces are considered simultaneously during landmark annotation.

Pairwise methods are more frequently used in the existing literature. The most prominent approach is based on fitting, or warping, a face template or a parametric surface model independently on each of the surfaces in the dataset. 3D Morphable Models \cite{blanz1999morphable}, Point Distribution Models \cite{claes2012improved, ghosh2009feature, hutton2004dense} and Nonrigid Iterative Closest Point \cite{amberg2007optimal} algorithms are primary examples of this category. Early literature work on these methodologies required an initial manual annotation of a number of landmarks in either a training set of surfaces, or even on the complete dataset. Subsequent extensions attempted to weaken or alleviate this requirement by incorporating the use of additional optimization criterions, such as the minimum description length principle or the maximization of mutual information, see for example  \cite{thodberg2003minimum, rueckert1999nonrigid, guo2013automatic, kakadiaris2007three}. Another approach in pairwise landmarking that has risen to attention in recent years entails the use of harmonic or conformal maps which project 3D meshes to a planar domain, thus transforming the 3D correspondence problem to one of 2D image matching \cite{wang2007conformal, wang2008high}. An initial annotation of a sparse set of correspondences is still a requirement here, and semi-automated methods have been proposed to achieve that, using for example alignment of mesh vertices through curvature features.

The second class of dense landmarking methodologies attempts to optimize the location of landmark points jointly on all considered surfaces. Although the result is the same as  applying one-to-one correspondences between all surfaces in a group and a template or mean surface model, considering all data at once can have certain advantages, such as removing the need for the construction of a template and reducing the bias that can result by the multiple pairwise fits to one surface \cite{van2011survey}. Groupwise functional correspondence methods reformulate the landmarking problem as one of finding correspondence between real-valued functional representations of the mesh surfaces, extracted for example using Heat or Wave Kernel Signatures \cite{ovsjanikov2012functional, zhang2016functional, bronstein2008numerical}. 

In the work presented by Cates et al. \cite{cates2007shape}, the groupwise correspondence was achieved by optimizing the compactness of the surfaces' distribution, represented as vectors of their landmark coordinates, subject to constraints enforcing uniform distribution of landmarks on individual surfaces. The method was based upon previous work on statistical shape modeling with information-based optimization functions, first presented by Kotcheff and Taylor \cite{kotcheff1998automatic}, and subsequent articles by Davies et al. \cite{davies2002minimum, davies20023d}. In contrast to the older methods, the algorithm by Cates et al. was not tailored towards the construction of a parametric surface model, but rather to directly solve the landmarking problem. Furthermore,  it did not necessitate the use of an anchor surface. Further details and comparisons between there methods can be found in Cates' PhD dissertation entitled 'Shape Modeling and Analysis with Entropy-Based Particle Systems'. 

Finally, a number of algorithms, usually coined as groupwise methodologies, adopt an intermediate approach to the correspondence problem. They are based on bottom-up iterative pairwise alignments of similar surfaces, driven by an affinity graph connecting similar surfaces \cite{tang2013image, wang2010groupwise}.

While extensive work has been done in the problem of identifying corresponding landmark points on sets of surfaces, as outlined previously, it is possible to identify some key methodological issues that are still prominent. In pairwise correspondence methods, the construction of a face model, or template, can be a tedious and error-prone procedure. Furthermore, the geometry of the final data objects after correspondence optimization, can be biased towards the mean or template surface. Iterative pairwise methods could address, to a certain extent, such problems, but have not been extensively used so far, probably due the extra computational cost they incur, as well as  the difficulty in constructing meaningful affinities between unregistered surfaces. Functional groupwise methodologies, on the other hand, may not suffer from the above issues, but are also associated with specific shortcomings. The most prominent, regarding the problem of point-to-point correspondence, being that the reverse mapping, from corresponding functions, to surface points, is not always easy to construct. 

In Cates et al. \cite{cates2007shape}, groupwise correspondence is achieved by manipulating the location of landmark points on the surfaces, such that an objective function comprised of two entropy-based terms is optimized. These terms are related to the uniform distribution of points in each surface and the overall landmark correspondence among surfaces. Due to the fact that landmark annotation under this formulation is equivalent to randomly sampling corresponding points from uniform distributions defined on the surfaces, the problem is also coined with the term ensemble surface sampling. The entropy formulation of the correspondence problem, and the associated optimization procedure, are attractive for a number of reasons. First, the methodology does not necessitate the construction of a template or any manual annotation. Second, the number of computed landmarks can be easily adjusted to the specific application requirements. Third, optimization can be easily tailored to problems of adaptive sampling, as will be discussed in the next section. However, two key problems of the previously presented method could be easily pinpointed. The optimization of the uniform distribution of points on individual surfaces was controlled by a kernel density estimator that did not take into consideration the structure of the surfaces. As a result, the method does not provide an optimal uniform distribution of landmarks and can not deal with highly curved surfaces. In addition, the gradient-descent optimization algorithm also ignored surface constraints. Point updates had to be recast on the surface after each iteration, which further impairs efficiency and performance.

In this work, we extend the methodology of Cates et al. \cite{cates2007shape} and present our Geodesic Ensemble Surface Sampling Algorithm (GESSA) for the automated identification of landmarks across sets of similar polyhedral surfaces. We propose a suitable estimator for the probability density function of a variable defined on a manifold or polyhedral surface, and employ it in the construction of the objective function. Furthermore, a gradient descent algorithm is constructed, which enables the optimization to be performed directly on the surfaces. incorporating these two key features inside the existing framework, we are able to to deal with highly curved surfaces and improve upon computational space requirements.

\subsubsection{Sparse Principal Component Analysis}

Different sparse PCA methods have been presented in the literature \cite{zou2006sparse, witten2009penalized, jolliffe2003modified}. Here use Penalized Matrix Decomposition (PMD), as proposed in \cite{witten2009penalized}, which has been shown to be similar but more computationally efficient than the SCoTLASS sPCA formulation \cite{witten2009penalized, jolliffe2003modified}.

Without loss of generality, let $\boldsymbol{P}$ be a column-wise zero-mean $N \times M$ data matrix. Standard PCA seeks unit loading vectors $\boldsymbol{v}_k$ so that linear transformations - principal components - $\boldsymbol{P} \boldsymbol{v}_k$ have successively maximum variance. The first PC loading vector is thus computed as
\begin{equation}\label{pca}
\boldsymbol{v}_1 = \argmax_v \boldsymbol{v}^T \boldsymbol{P}^T \boldsymbol{P} \boldsymbol{v},\ s.t. \  \boldsymbol{v}^T \boldsymbol{v} = 1
\end{equation}
Consequent loading vectors can be computed by repeating the same process on the deflated data matrices. Given $\boldsymbol{P}_k$ and $\boldsymbol{v}_k$, the deflated data matrix $\boldsymbol{P}_{k+1} = \boldsymbol{P}_k - \boldsymbol{P}_k \boldsymbol{v}_k \boldsymbol{v}_k^T$, with  $\boldsymbol{P}_1 = \boldsymbol{P}$.

The SCoTLASS procedure for sPCA modifies the optimization problem \eqref{pca} with an additional $L_1$ regularization constraint on the loading vectors: $\| \boldsymbol{v} \|_1 \leq t$, for some tuning parameter $t$. It has been shown that for the first PC, SCoTLASS is equivalent to the following penalized matrix decomposition problem \cite{witten2009penalized}:
\begin{equation}
\boldsymbol{v}_1 = \argmax_{\boldsymbol{u},\boldsymbol{v}} \boldsymbol{u}^T\boldsymbol{P} \boldsymbol{v},\ s.t. \ \| \boldsymbol{v} \|_1 \leq s_p,\  \| \boldsymbol{v}\|_2^2 \leq 1,\  \| \boldsymbol{u} \|_2^2 \leq 1,
\end{equation}
where $s_p$ is the sparsity parameter, with lower values leading to sparser loading vectors $\boldsymbol{v}$. The above problem is biconvex and can be optimized by iteratively alternating between maximization with respect to $\boldsymbol{u}$ and $\boldsymbol{v}$. SCoTLASS imposes orthogonality constraints between subsequent loading vectors, which though makes optimization very difficult. PMD does not utilize such constraints. Consequent components in PMD are again computed by applying the same procedure on the deflated data matrices. 

We notice here that sPCA does not guarantee uncorrelated principal components. We have opted to use sPCA since a main objective in our decomposition analysis was to construct composite traits corresponding to spatially coherent facial areas, which could not be achieved through standard PCA. This coherency was imposed by controlling the sparsity parameter $s_p$. Tuning the parameter is commonly performed through cross validation, by selecting the value that leads to minimum average CV reconstruction error of the data \cite{witten2013package}. This process though could have led to spatially extended loading vectors for which biological interpretation would be difficult. Furthermore, the parameter would need tuning for each principal component independently, which would be computationally expensive. 

Since our primary objective was to estimate the heritability of principal components, we evaluated the effect of sparsity by comparing heritability estimates (see below for detailed description) of a fixed number of PCs derived from different $s_p$ values. In detail, we computed $100$ sPCs for each curvature descriptor and $7$ different parameter values. Heritability estimates for all components were subsequently computed. We noticed that the sorted heritability estimates show similar behavior across curvature descriptors for all $s_p$ values tested.

Based on the fact that heritability estimation would not be significantly affected by the particular value of $s_p$, we selected constant values for each descriptor after visual inspection of sPC loading maps - constructed by mapping the sPCs' loadings on the facial surface - with the criterion of which parameter yielded sPCs more suitable for further biological interpretation. In particular, the parameter was set to $7.5$ for GC, $12.5$ for SI and $15$ for MC and CU. 

We retained for further analysis all $100$ sPCs for each curvature descriptor. Each set of sPCs was able to cumulatively explain respectively $92.72\%$, $89.41\%$, $90.11\%$ and $88.2\%$ of the MC, GC, CU and SI phenotypic curvature variance.

\subsubsection{Structural Equation Modelling}

In this work, we estimated heritability as the proportion of phenotypic variance explained by genetic factors. Since the genetic and environmental variables are unobserved (latent), their effects are inferred from twin resemblance using Structural Equation Modelling (SEM). SEM encompasses a broad family of statistical modeling techniques and can be viewed as a combination of path and latent factor analysis \cite{hox11introduction}. SEM is also widely referred to as covariance structure modeling, since a SE model implies a structure for the covariances between observed variables.

\begin{figure}[ht]
\centerline{\includegraphics[width=3.5in]{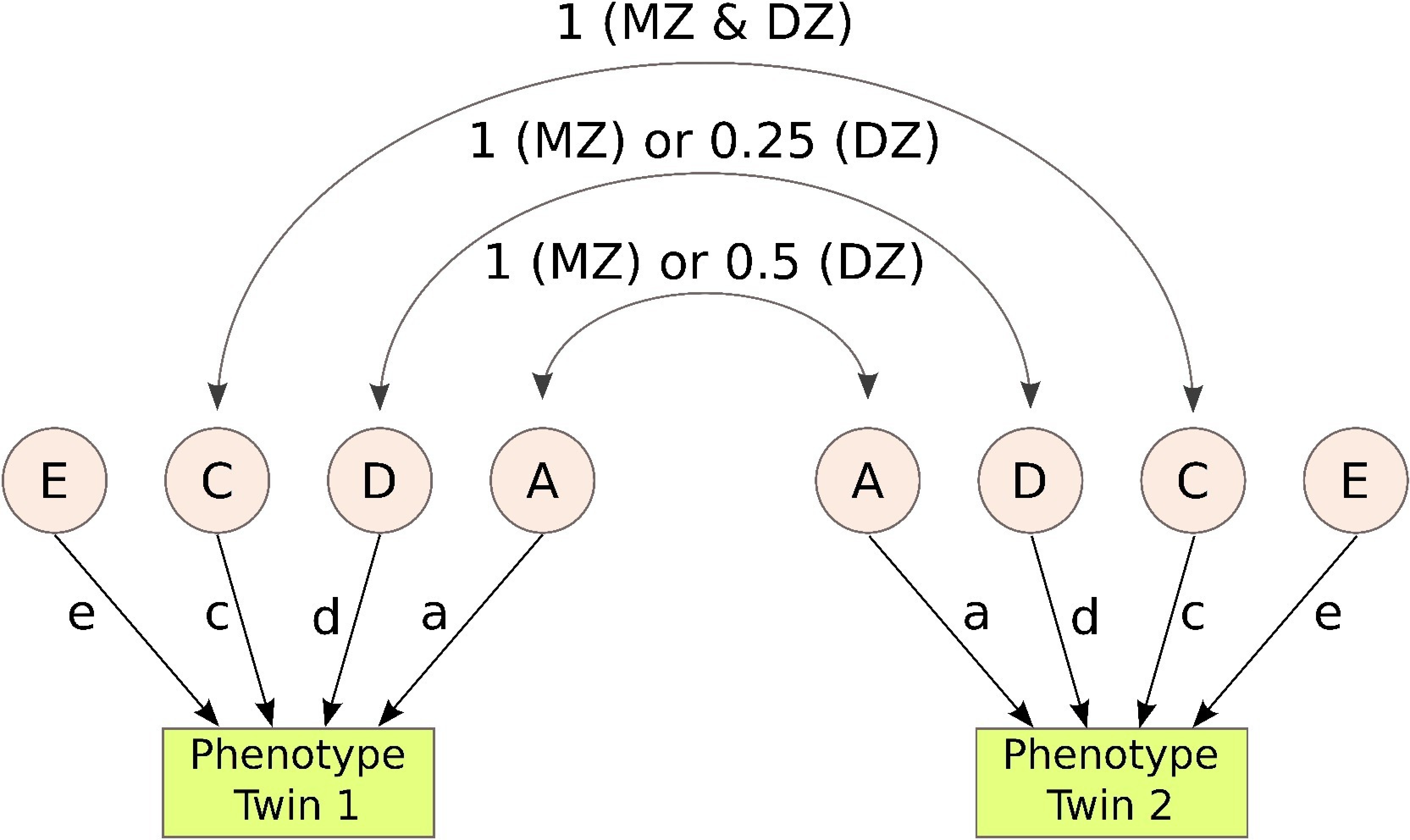}}
\caption{{\bf Path Diagram of the ACDE Structural Equation Model.} Latent factors $A$, $D$, $C$ and $E$ correspond to additive and dominant genetic, common and unique environmental effects respectively. Double arrows represent the latent factor correlations between pairs of twins. Additive and dominant genetic effects are correlated $1$ between MZ pairs, $0.5$ and $0.25$ between DZ pairs. Common environmental effects have correlation $1$ for both types of twins while unique environmental factors are uncorrelated. $a$, $d$, $c$ and $e$ are regression path coefficients of the respective latent factors. Heritability in a ACDE SEM model is given by $h_{ACDE}^2 = \frac{a^2 + d^2}{a^2 + d^2 + c^2 + e^2}$.}
\label{fig_S9}
\end{figure}

In heritability studies, observed phenotypic variation can be partitioned into variance components from the following latent factors: additive (A) genetic, dominant (D) genetic, common (C) environmental and unique (E) environmental, with the latter component also including measurement error. A path diagram of a SEM including all of the above latent factors can be seen in Figure \ref{fig_S9}. The structure of covariance is implied by latent factor correlations between twin pairs. Additive and dominant genetic effects are correlated $1$ between MZ pairs, while only $0.5$ and $0.25$ respectively between DZ pairs. Common environmental effects have correlation $1$ for both types of twins while unique environmental factors are uncorrelated.

For twin studies in particular, the model ADCE is over-specified and cannot be estimated using twin data alone \cite{rijsdijk2002analytic}. Submodels that can be examined are ACE, ADE, AE, CE and E. Models composed of the components DE are not considered biologically plausible. Here we considered ACE, AE and E models. As such, below we focus and describe in detail the definition of the ACE SE model.

A univariate ACE model can be expressed as
\begin{equation}
P = aA + cC + eE,
\end{equation}
where for simplicity and without loss of information, $P$ is an observed zero-centered, continuous phenotypic variable and $A$, $C$, $E$ are unobserved latent factors with fixed unit variances and covariances that depend on the type of twin. Finally, $a$, $c$, $e$ are regression coefficients expressing the effects of the latent variables in the phenotype. Now Let $\boldsymbol{P}_{MZ}$ be a $N_{MZ} \times 2$ matrix of phenotypic observations with each row coming from one pair of MZ twins and $N_{MZ}$ the number of MZ paired observations. Respectively $\boldsymbol{P}_{DZ}$ denotes a $N_{DZ} \times 2$ matrix of phenotypic observations with each row coming from one pair of DZ twins. Furthermore let 
\begin{equation}
\boldsymbol{L} = \begin{bmatrix}
       a & 0           \\
       c & 0           \\
       e & 0           \\
       0 & a           \\
       0 & c           \\
       0 & e           \\
      
     \end{bmatrix},
\end{equation}
be the matrix of regression coefficients and finally $\boldsymbol{\Lambda}_{MZ}$, $\boldsymbol{\Lambda}_{DZ}$ be $N_{MZ} \times 6$ and $N_{DZ} \times 6$ matrices of unobserved A, C, E factors for MZ and DZ twin subsets respectively.

Expressing the ACE model for our observations, we have
\begin{equation}
\boldsymbol{P}_{MZ} = \boldsymbol{\Lambda}_{MZ} \boldsymbol{L}, \ \boldsymbol{P}_{DZ} = \boldsymbol{\Lambda}_{DZ} \boldsymbol{L}.
\end{equation}

The expected phenotypic covariances from the ACE model are:
\begin{equation}
\begin{split}
\Sigma_{MZ} & = E[ \boldsymbol{\Lambda}_{MZ}^T \boldsymbol{\Lambda}_{MZ}] = \boldsymbol{L}^T \boldsymbol{\Psi}_{MZ}\boldsymbol{L} \\
\Sigma_{DZ} & = E[ \boldsymbol{\Lambda}_{DZ}^T \boldsymbol{\Lambda}_{DZ}] = \boldsymbol{L}^T \boldsymbol{\Psi}_{DZ}\boldsymbol{L},
\end{split}
\end{equation}
where the correlation matrices $\boldsymbol{\Psi}_{MZ}$, $\boldsymbol{\Psi}_{DZ}$ of the latent factors are derived from the SEM path diagram. In particular, $\boldsymbol{\Psi}_{MZ}$ has two off diagonal elements equal to $1$, corresponding to $corr(A_1, A_2)$ and $corr(C_1, C_2)$, while $\boldsymbol{\Psi}_{DZ}$ has  $corr(A_1, A_2) = 0.5$ and $corr(C_1, C_2) = 1$ (see Figure \ref{fig_S9}).

The structured covariance matrices as modelled by SEM can be now easily computed to be
\begin{equation}
\Sigma_{MZ} = \begin{bmatrix}
       a^2 + c^2 + e^2 & a^2 + c^2          \\
       a^2 + c^2 & a^2 + c^2 + e^2           \\
     \end{bmatrix},
\end{equation}
\begin{equation}
\Sigma_{DZ} = \begin{bmatrix}
       a^2 + c^2 + e^2 & 0.5a^2 + c^2          \\
       0.5a^2 + c^2 & a^2 + c^2 + e^2           \\
     \end{bmatrix},
\end{equation}

Maximum Likelihood is used to estimate the regression coefficients  $a$, $c$, $e$. Let $S_{MZ}$ and $S_{DZ}$ be the observed sample covariances. Assuming that the phenotypic response is normally distributed, the probabilities of observing $S_{MZ}$ and $S_{DZ}$ given estimates $\hat{\Sigma}_{MZ}$ and $\hat{\Sigma}_{DZ}$ follow the Wishart distribution with $N_{MZ}$ and $N_{DZ}$ degrees of freedom respectively. The  log-likelihood functions can be written as follows, after the omission of constant terms \cite{martin1977genetical}:
\begin{equation}
\begin{split}
-2LL_{MZ} & \approx N_{MZ} \left[ \ln | \hat{\Sigma}_{MZ}  | +  tr(\hat{\Sigma}_{MZ}^{-1} S_{MZ})   \right] \\
-2LL_{DZ} & \approx N_{DZ} \left[ \ln | \hat{\Sigma}_{DZ}  | +  tr(\hat{\Sigma}_{DZ}^{-1} S_{DZ})   \right].
\end{split}
\end{equation}

Estimates $\hat{a}, \hat{c}, \hat{e}$ are obtained by maximizing the combined likelihood function $-2 (LL_{MZ} + LL_{DZ})$. Model fit can be assessed using a log-likelihood ratio test between the structured model and a fitted saturated model where no structure is imposed on the covariances. The ratio statistic is distributed approximately as a  chi-squared distribution with degrees of freedom equal the difference in $df$ between the structured and the saturated model.

At this point we can also define the heritability estimate $h_{ACE}^2$ from the ACE model as
\begin{equation}
h_{ACE}^2 = \frac{\hat{a}^2}{\hat{a}^2 + \hat{c}^2 + \hat{e}^2}.
\end{equation}

An important aspect of SEM in twin studies is that the significance of individual variance components can be assessed by dropping parameters sequentially from nested models; here ACE$\rightarrow$AE$\rightarrow$E. In choosing between models, variance components are excluded in the selection process if there is no significant deterioration in model fit, assessed commonly by the Akaike Information Criterion (AIC) \cite{akaike1974new}, after the component is dropped. The E component represents random error and is always retained \cite{rijsdijk2002analytic}. Heritability is estimated from the AE model as
\begin{equation}
h_{AE}^2 = \frac{\hat{a}^2}{\hat{a}^2 + \hat{e}^2}.
\end{equation}

In this study, we estimated heritability for all $4,096$ curvature traits independently, as well as for the top $100$ variance-explaining sPCs (composite traits), for each curvature descriptor, using SEM. We assessed the significance of individual variance components by dropping parameters sequentially from the set of nested models ACE, AE and E, fitted using the OpenMx software \cite{boker2011openmx,url_mx}. Age was included in the models as a covariate. 

\FloatBarrier

\end{document}